\documentclass{elsart}
\usepackage{graphicx}
\usepackage[fleqn]{amsmath}
\usepackage{amssymb}
\catcode`\@=11

\@addtoreset{equation}{section}

\newcommand{\be}{\begin{equation}}\newcommand{\ee}{\end{equation}}
\newcommand{\bea}{\begin{eqnarray}}\newcommand{\eea}{\end{eqnarray}}

\newcommand{\F}{\phantom {1}}
\newcommand{\Fm}{$\phantom {-}$}

\newcommand{\<}{\langle}\renewcommand{\>}{\rangle}
\newcommand{\nn}{\nonumber}
\renewcommand{\[}{\langle\!\langle}
\renewcommand{\]}{\rangle\!\rangle}
\newcommand{\itsl}{i_{\textrm{tsl}}}
\newcommand{\jtsl}{j_{\textrm{tsl}}}
\newcommand{\Ntsl}{N_{\textrm{tsl}}}
\newcommand{\isub}{i_{\textrm{sub}}}
\newcommand{\Nsub}{N_{\textrm{sub}}}
\newcommand{\ired}{i_{\textrm{red}}}
\newcommand{\Tr}{\textrm{Tr}}
\newcommand{\Niupd}{N_{\textrm{iupd}}}

\allowdisplaybreaks[2]

\begin{document}

\begin{frontmatter}
\rightline{\small DESY 06-061, MKPH-T-06-11, RCNP-Th 06003}
\title{Spin-dependent potentials from lattice QCD}

\author{Yoshiaki Koma$^{a,b}$ and Miho Koma$^{a,b,c}$}

\address{$^a$Deutsches Elektronen-Synchrotron DESY, Theory Group, 
22607 Hamburg, Germany\\
$^b$Institut f\"ur Kernphysik,
Johannes Gutenberg-Universit\"at Mainz, 55099 
Mainz, Germany\\
$^c$Research Center for Nuclear Physics (RCNP), Osaka University,
Osaka 576-0047, Japan}

\begin{abstract}
The spin-dependent corrections to the static inter-quark
potential are phenomenologically relevant to describing
the fine and hyperfine spin splitting of the heavy
quarkonium spectra.
We investigate these corrections, which are represented as
the field strength correlators on the quark-antiquark source,
in SU(3) lattice gauge theory.
We use the Polyakov loop correlation function as the
quark-antiquark source, and by employing the multi-level algorithm,
we obtain remarkably clean signals for these corrections up to
intermediate distances of around 0.6 fm.
Our observation suggests several new features of the corrections.
\end{abstract}

 
\end{frontmatter}

\section{Introduction}
\label{sec:intro}
\vspace*{-0.1cm}

\par
The spin-dependent potentials are parts of relativistic 
corrections to the static quark-antiquark potential,
which depend on quark spin, and
are phenomenologically relevant to describing 
the fine and hyperfine splitting of heavy quarkonium 
spectra~\cite{Lucha:1991vn,Buchmuller:1992zf,%
Bali:2000gf,Brambilla:2004wf}.
Thus it is interesting to address these 
corrections from QCD and to compare with 
the observed spectra.

\par
The relativistic corrections are usually classified 
in powers of the inverse of heavy quark mass $m$
(or quark velocity $v$)
and it is well-known that in QCD
the leading spin-dependent corrections show up
at $O(1/m^{2})$~\cite{Eichten:1979pu,Eichten:1980mw,%
Peskin:1983up,Gromes:1983pm,Ng:1985uq,Pantaleone:1985uf}.
These spin-dependent corrections
were also derived systematically within 
an effective field theory framework called
potential nonrelativistic QCD 
(pNRQCD)~\cite{Pineda:2000sz}.
pNRQCD is obtained by integrating out the scales 
above $m\gg \Lambda_{\rm QCD}$ in 
QCD~\footnote{$\Lambda_{\rm QCD}$ is assumed to be a 
few hundred of MeV} first,
which leads to NRQCD~\cite{Caswell:1985ui,Bodwin:1994jh},
and then $mv$, leaving a typical scale of the 
binding energy of heavy quarkonium 
$m v^2$~\cite{Brambilla:1999xf,Brambilla:2000gk,Brambilla:2004jw}.

\par
The spin-dependent potential is summarized in the form
\bea
V_{\rm SD}(r) &=& 
\left (
\frac{\vec{l}_{1} \cdot\vec{s}_{1} }{m_{1}^{2}}
- \frac{\vec{l}_{2}\cdot\vec{s}_{2} }{m_{2}^{2}}
\right )
\left ( 
\frac{(2c_{F}^{(+)}-1)V_{0}'(r)+ 2 c_{F}^{(+)} V_{1}'(r)}{2r}
\right )
\nonumber\\*
&&
+
\left (
\frac{\vec{l}_{1} \cdot\vec{s}_{2} }{m_{1}m_{2}}
-\frac{\vec{l}_{2}\cdot\vec{s}_{1}  }{m_{1}m_{2}}  
\right )
\frac{c_{F}^{(+)} V_{2}'(r)}{r}
\nonumber\\*
&&
+\left (
\frac{\vec{l}_{1}\cdot \vec{s}_{1} }{m_{1}^{2}}
+ \frac{\vec{l}_{2} \cdot\vec{s}_{2} }{m_{2}^{2}}
\right )
\left ( \frac{ c_{F}^{(-)} ( V_{0}'(r)+ V_{1}'(r))}{r}
\right )
\nonumber\\*
&&
+
\left (
\frac{\vec{l}_{1}\cdot\vec{s}_{2} }{m_{1}m_{2}}
+\frac{ \vec{l}_{2} \cdot\vec{s}_{1}}{m_{1}m_{2}}  
\right )
\frac{c_{F}^{(-)} V_{2}'(r)}{r}
\nonumber\\*
&&
 +
 \frac{1}{m_{1}m_{2}}
 \left ( 
 \frac{(\vec{s}_{1}\cdot \vec{r})( \vec{s}_{2}\cdot \vec{r})}
 {r^{2}} -\frac{\vec{s}_{1}\cdot \vec{s}_{2}}{3} 
 \right ) 
 c_{F}^{(1)} c_{F}^{(2)}
 V_{3}(r)
 \nonumber\\*
 &&
 +
 \frac{\vec{s}_{1}\cdot \vec{s}_{2}}{3m_{1}m_{2}}
 \left (
 c_{F}^{(1)}c_{F}^{(2)}V_{4}(r) -48 \pi C_{F} \alpha_{s} d_{v} 
 \delta^{(3)} (r) 
\right )\; ,
\label{eqn:potential}
\eea
where $\vec{r}_{1}$ and $\vec{r}_{2}$ 
($r \equiv |\vec{r}_{1}-\vec{r}_{2}|$)
denote the positions of quark and antiquark,
$m_{1}$ and $m_{2}$ the masses,
$\vec{s}_{1}$ and $\vec{s}_{2}$ the 
spins ($\vec{s} = \vec{\sigma}/2$ with $\vec{\sigma}$
being the Pauli matrices),  and
$\vec{l}_{1} = -\vec{l}_{2} =\vec{l}$ 
the orbital angular momenta.
$V_{0}(r)$ is the spin-independent static potential
at $O(m^0)$ and the prime denotes the derivative
with respect to~$r$.
$V_{1}'(r)$, $V_{2}'(r)$, $V_{3}(r)$ and $V_{4}(r)$ 
are the functions which depend only on~$r$.
In what follows we call these functions loosely the 
spin-dependent potentials.
$c_{F}^{(i)}(\mu, m_{i})$ ($i=1,2$) is 
the matching coefficient in the (p)NRQCD
Lagrangian which multiplies the term 
$\vec{\sigma}\cdot\vec{B}/(2m_{i})$ and this
coefficient plays an important role when connecting
QCD at a scale $\mu$ with (p)NRQCD at scales~$m_{i}$.
We have defined as $c_{F}^{(\pm)}=
(c_{F}^{(1)} \pm c_{F}^{(2)})/2$.
For equal quark and antiquark masses ($m_{1}=m_{2}$), 
$c_{F}^{(-)}$ vanishes as $c_{F}^{(1)}=c_{F}^{(2)}$.
When the matching is performed at 
tree-level of perturbation theory,
the coefficient is $c_{F}^{(i)}=1$~\cite{Manohar:1997qy}
and then Eq.~\eqref{eqn:potential} is
reduced to the expression given 
in Refs.~\cite{Eichten:1979pu,Eichten:1980mw,Gromes:1983pm}.
$\alpha_{s}=g^2/(4\pi)$ is the strong coupling and
$C_{F}=4/3$ the Casimir charge of the fundamental
representation, and $d_{v}$ the mixing coefficient 
of the four-quark operator in the (p)NRQCD 
Lagrangian (see e.g. 
the Appendix~E of Ref.~\cite{Bali:2000gf}).

\par
Given the field strength $F_{\mu\nu}$, where
the color-electric and the color-magnetic fields are defined by
$E_{i}=F_{4i}$ and
$B_{i}=\epsilon_{ijk} F_{jk}/2$, 
respectively,\footnote{
Throughout this paper we work in 
Euclidean space and assume that 
the repeated spinor (Latin) and color (Greek)
indices are summed over from 1 to 4 and from 1 to 3,
respectively, unless explicitly stated.}
the spin-dependent potentials in Eq.~\eqref{eqn:potential}
are expressed as
\bea
&&
\frac{r_{k}}{r}
V_{1}'(r)= \epsilon_{ijk}
\lim_{\tau \to \infty}\int_{0}^{\tau} dt \; t
\[ g^2 B_{i}(\vec{r}_{1}, t_{1}) E_{j}(\vec{r}_{1},t_{2}) \]  \; , 
\label{eqn:pot1}\\
&&
\frac{r_{k}}{r}
V_{2}'(r) = \epsilon_{ijk}\lim_{\tau \to \infty}\int_{0}^{\tau} d t \; t
\[  g^2  B_{i}(\vec{r}_{1}, t_{1}) E_{j}(\vec{r}_{2},t_{2}) \]  \; ,  
\label{eqn:pot2}\\
&&
(\frac{r_{i}r_{j}}{r^2}-\frac{\delta_{ij}}{3})
V_{3}(r)
+\frac{\delta_{ij}}{3}V_{4}(r)
=
2\lim_{\tau \to \infty}\int_{0}^{\tau} dt \;
\[ g^2  B_{i}(\vec{r}_{1}, t_{1})B_{j}(\vec{r}_{2},t_{2}) \] \; .
\label{eqn:pot3-4}
\eea
Here $t\equiv t_{2}-t_{1}$ denotes the relative temporal 
distance between two field strength operators.
The double bracket $\[ \cdots \]$ represents 
the expectation value of the field strength correlator,
where the field strength operators
are attached to the quark-antiquark
source in a gauge invariant way.
In Refs.~\cite{Eichten:1979pu,Eichten:1980mw,Gromes:1983pm},
these expressions were given in the double-integral
form with the Wilson loop, 
which can be reduced to the single-integral form
through the spectral representation of the field strength
correlators derived from the transfer matrix theory.
However, it should be noted that 
the authors of Ref.~\cite{Pineda:2000sz}
pointed out that one of the spin-orbit potentials
$V_{2}'(r)$ in Refs.~\cite{Eichten:1979pu,Eichten:1980mw,Gromes:1983pm}
was underestimated by a factor two.
The expressions in Eqs.~\eqref{eqn:pot1}-\eqref{eqn:pot3-4}
are consistent with Ref.~\cite{Pineda:2000sz}
apart from the space-time metric; here we employ
the Euclidean metric, while
the Minkowski metric is used in Ref.~\cite{Pineda:2000sz}.%
~\footnote{The change of metric from 
Minkowski to Euclidean space-time is achieved by
$t^{(M)} \to -it^{(E)}$, $E^{(M)} 
\to i E^{(E)}$,  $B^{(M)}  \to B^{(E)}$.}

\par
As the expressions of the spin-dependent
potentials in Eqs.~\eqref{eqn:pot1}-\eqref{eqn:pot3-4}
are essentially nonperturbative, these potentials can be
studied in a framework beyond perturbation theory,
for instance, by using lattice QCD Monte Carlo simulations.
Rather, as the typical scale of the momentum 
$mv$ can be of the order as $\Lambda_{\rm QCD}$ due to 
nonrelativistic nature of the system $v\ll 1$, it is not
clear {\it a priori} that the perturbative determination of 
the potential is justified, and
indeed, nonperturbative contributions are expected in
the spin-orbit potentials $V_{1}'(r)$ and $V_{2}'(r)$;
they are related to the static potential through the Gromes 
relation~\cite{Gromes:1984ma,Brambilla:2001xk}, i.e.
$V_{0}'(r)=V_{2}'(r)-V_{1}'(r)$, where
$V_{0}(r)$ is known to contain
a nonperturbative long-ranged component
characterized by the string tension.
This relation was derived by exploiting 
the Lorentz invariance of the field 
strength correlators, which does not depend
on the order of perturbation theory.
 
\par
The determination of the spin-dependent  potentials using
lattice QCD simulations goes back
to the 1980s~\cite{deForcrand:1985zc,%
Michael:1985wf,Michael:1985rh,Huntley:1986de,%
Campostrini:1986ki,Campostrini:1987hu,%
Ford:1988as,Koike:1989jf} 
and to the 1990s~\cite{Born:1993cp,Bali:1996cj,Bali:1997am}.
The qualitative findings (quantitative to some extent) in these
earlier studies indicated that the spin-orbit 
potential $V_{1}'(r)$ contains 
the long-ranged nonperturbative component, while
all other potentials are relevant only to short-range physics
as expected from the one-gluon exchange interaction.
However, one observes that the spin-dependent potentials
(in particular, the spin-orbit potential $V_{1}'$)
of even the latest studies~\cite{Bali:1996cj,Bali:1997am}
suffer from large numerical errors, which can obscure
their behavior already at intermediate distances.
For phenomenological applications of these potentials,
it is clearly important to determine their functional form
as accurately as possible.

\par
In the present paper we thus revisit the determination
of the spin-dependent potentials with lattice QCD within
the quenched approximation, 
aiming at reducing the numerical errors with a new approach.
There are mainly two possible sources of numerical
errors apart from the systematic error due to 
discretization of space-time.
One is the statistical error for the expectation value of the 
field strength correlator, and the other is the systematic
error associated with the integration over $\tau$ and the 
extrapolation of $\tau \to \infty$ in 
Eqs.~\eqref{eqn:pot1}-\eqref{eqn:pot3-4}.
In order to control   the total error, first of all,
one needs to evaluate the field strength correlator 
precisely, as otherwise its uncertainty is enhanced in the 
following procedures.

\par
Our idea is then to employ the 
multi-level algorithm~\cite{Luscher:2001up,Luscher:2002qv}
for measuring the field strength 
correlators~\cite{Koma:2005nq,Koma:2006si},
as we expect clean signals even at larger~$r$ and~$t$.
This algorithm also allows us to use the Polyakov loop
correlation function (PLCF), a pair of Polyakov loops $P$ 
separated by a distance $r$, as the quark-antiquark
source instead of the Wilson loop.
In fact, if one relies on the commonly employed simulation algorithms,
it is almost impossible to evaluate the field strength correlators
on the PLCF, or the PLCF itself, at intermediate distances
with reasonable computational effort, since
the expectation value of the PLCF at zero temperature
is smaller by several orders of magnitude
than that of the Wilson loops and is easily hidden in  
the statistical noise.
However, as we will show in the next section, if one manages to
obtain accurate data for the field strength correlators on the PLCF,
the systematic errors from the integration and the extrapolation can
be avoided. The key idea is to employ the spectral representation of
the field strength correlators on the PLCF, which is plugged into
Eqs.~\eqref{eqn:pot1}-\eqref{eqn:pot3-4}.

\par
This paper is organized as follows.
In sect.~\ref{sec:procedure}, we 
describe our procedures to obtain the spin-dependent
potentials, which contain the derivation of the 
spectral representation of the
field strength correlators on the PLCF and of the
spin-dependent potentials, the definition of the 
field strength operators on the lattice, and the
implementation of the multi-level algorithm.
In sect.~\ref{sec:result}, we show numerical results, followed by
analyses and discussions.
The summary is given in sect.~\ref{sec:summary}.
In this paper, we will not discuss the matching coefficient
but the interested reader can refer to 
the discussion in Ref.~\cite{Bali:2000gf}.
We also plan to revisit this issue in our future studies.

\section{Numerical procedures} 
\label{sec:procedure}

\par
In this section, we describe the spectral representation of the
field strength correlators on the PLCF and of the 
spin-dependent potentials. We provide the definition of the 
field strength operators on the lattice, and explain the
implementation of the multi-level algorithm.
The standard Wilson action is most preferable
for this algorithm because its action 
density is locally defined by plaquette and 
thus we shall use this action in our present simulation.
The lattice volume is $L^3 T$ and
periodic boundary conditions are imposed
in all directions.

\subsection{Spectral representation of the 
field strength correlators and of the spin-dependent potentials}
\label{subsect:spectral-rep}

\par
Let us derive the spectral representation of the field strength
correlators on the PLCF using the transfer matrix formalism.
We follow the notation used in Ref.~\cite{Luscher:2002qv}, 
in which the spectral representation of the PLCF is discussed.
We consider the transfer matrix in the 
temporal gauge $\mathbb{T}\equiv e^{-\mathbb{H}a}$
which acts on the states on the space of all spatial lattice 
gauge fields $U_{\mu}$ at a given time, where $a$ denotes
the lattice spacing.
We also introduce the projectors
$\mathbb{P}$ onto the subspace of gauge-invariant states and
$\mathbb{P}_{{\bf 3}\otimes {\bf 3}^{*}}(\vec{r}_{1},\vec{r}_{2})$ 
to the subspace of the states in the 
${\bf 3}\otimes {\bf 3}^{*}$ representation of SU(3).
Then the partition functions in the sector corresponding to
$\mathbb{P}$ and 
$\mathbb{P}_{{\bf 3}\otimes {\bf 3}^{*}}(\vec{r}_{1},\vec{r}_{2})$
are given by
$\mathcal{Z}=\Tr \{\mathbb{P}\,  e^{-\mathbb{H}T} \}$ and 
$\mathcal{Z}_{{\bf 3}\otimes 
{\bf 3}^{*}}(\vec{r}_{1},\vec{r}_{2})  \equiv 
\frac{1}{9} \Tr \{\mathbb{P}_{{\bf 3}\otimes {\bf 3}^{*}}
(\vec{r}_{1},\vec{r}_{2}) e^{-\mathbb{H}T} \}$, respectively.

\par
Firstly, we consider the
spectral representation of a double-bracket correlator
for operators $O_{1}(t_{1})$ and  $O_{2}(t_{2})$, which
are attached to either side of the PLCF
(the same side or the opposite side),
\bea
\[O_{1}(t_{1})O_{2}(t_{2}) \] &\equiv&
\frac{\<  O_{1}(t_{1}) O_{2}(t_{2}) 
\>_{P(\vec{r}_{1})P^*(\vec{r}_{2})}}
{\<P(\vec{r}_{1})P^*(\vec{r}_{2}) \>}
\nn\\*
&=&
\frac{ \frac{1}{9}\Tr 
\left[\mathbb{P}_{{\bf 3}\otimes
{\bf 3}^{*}}(\vec{r}_{1},\vec{r}_{2})
e^{-\mathbb{H}(T-t)} O_{1}
e^{-\mathbb{H}t} O_{2}\right]}{\mathcal{Z}}
\frac{\mathcal{Z}}
{\mathcal{Z}_{3\otimes 3^{*}}(\vec{r}_{1},\vec{r}_{2})} \nn\\*
&=&
\frac{ \frac{1}{9}\Tr 
\left[\mathbb{P}_{{\bf 3}\otimes
{\bf 3}^{*}}(\vec{r}_{1},\vec{r}_{2})
e^{-\mathbb{H}(T-t)} O_{1}
e^{-\mathbb{H}t} O_{2}\right]}{
\mathcal{Z}_{3\otimes 3^{*}}(\vec{r}_{1},\vec{r}_{2}) } \; ,
\eea
where we have used the identity
$\<P(\vec{r}_{1})P^{*}(\vec{r}_{2}) \>
=\mathcal{Z}_{{\bf 3}\otimes {\bf 3}^{*}}
(\vec{r}_{1},\vec{r}_{2})/\mathcal{Z}$.
Inserting the complete set of eigenstates
in the ${\bf 3}\otimes {\bf 3}^{*}$ representation
$|n \> \equiv |n; \vec{r}_{1},\vec{r}_{2}\>$,
which satisfy $\mathbb{T}
|n \> =e^{-E_{n}(r)a }|n\>$ with energies $E_{n}(r) > 0$, we obtain
\bea
\[ O_{1}(t_{1}) O_{2}(t_{2}) \]
=
\frac{\sum_{n= 0,m=0}^{\infty} \<n | O_{1}|m \>\<m | O_{2}| n\>
e^{-E_{m}t} e^{-E_{n}(T-t)} }{\sum_{n=0}^{\infty} e^{-E_{n}T}} \; .
\label{eqn:o1o2}
\eea
We denote the energy gap between two eigenstates
as $\Delta E_{mn}(r) = E_{m}(r) - E_{n}(r)$. 
Then, up to terms involving exponential factors equal 
to or smaller than $e^{-(\Delta E_{10})T}$,
Eq.~\eqref{eqn:o1o2} is reduced to
\bea
&&\[ O_{1}(t_{1}) O_{2}(t_{2}) \]
= 
\<0 | O_{1}|0 \> \<0| O_{2}| 0\> \nn\\*
&&
+\sum_{m = 1}^{\infty}
\biggl (\<0 | O_{1}|m \> \<m | O_{2}| 0\>
e^{-(\Delta E_{m0})t}  
+
 \<m | O_{1}|0 \> \<0 | O_{2}| m\>
 e^{-(\Delta E_{m0})(T-t)}\biggr)
\nn\\*
&& 
+O(e^{-(\Delta E_{10})T})\; .
\label{eqn:o1o2-final}
\eea

\par
In the case of the field strength correlators, 
we can further simplify Eq.~\eqref{eqn:o1o2-final}
by using the properties of the color-magnetic 
and color-electric field operators under the time reversal;
we have relations
\bea
&&
\[g^2 B_{i}(t_{1})E_{j}(t_{2}) \]
= -\[g^2  B_{i}(t_{2})E_{j}(t_{1}) \] \; ,\\
&&
\[g^2  B_{i}(t_{1})B_{j}(t_{2}) \]
= \[g^2 B_{i}(t_{2})B_{j}(t_{1}) \] \; ,
\eea
which, for the matrix elements, read
\bea
&&
\<m| gB_{i} |0\> \<0 | gE_{j}|m \> =
- \<0| gB_{i}|m\> \<m | gE_{j}|0 \> \; ,\\
&&\<m| gB_{i} |0\> \<0 | gB_{j}|m \> =
\<0| gB_{i}|m\> \<m | gB_{j}|0 \> \; ,
\eea
for $m \geq 1$.
Moreover, $\<0| gB_{i}|0\>=0$
since $B_{i}$ is odd under CP transformations.
The field strength correlators in 
Eqs.~\eqref{eqn:pot1}-\eqref{eqn:pot3-4}
are thus expressed as
\bea
\[ g^2 B_{i}(\vec{r}_{1}, t_{1}) E_{j}(\vec{r}_{1},t_{2}) \] 
&=&
2\sum_{m = 1}^{\infty}
\<0 | gB_{i}(\vec{r}_{1})|m \>
\<m | gE_{j}(\vec{r}_{1})| 0\> 
\nn\\*
&&\quad \times
e^{-(\Delta E_{m0})T/2} 
\sinh ((\Delta E_{m0})(T/2 -t)) \nn\\*
&&+O(e^{-(\Delta E_{10})T})\;,
\label{eqn:spectral-rep1}
\\
\[g^2  B_{i}(\vec{r}_{1}, t_{1}) E_{j}(\vec{r}_{2},t_{2}) \]
& =&
2\sum_{m = 1}^{\infty} 
\< 0 | gB_{i} (\vec{r}_{1}) | m\>
\<m |   gE_{j}(\vec{r}_{2}) | 0\>
\nn\\*
&&\quad\times
e^{-(\Delta E_{m0})T/2} 
\sinh ((\Delta E_{m0})(T/2 -t)) \nn\\*
&& +O(e^{-(\Delta E_{10})T})\;,
\label{eqn:spectral-rep2}
\\
\[g^2 B_{i}(\vec{r}_{1},t_{1})B_{j}(\vec{r}_{2},t_{2}) \]
&=& 
2\sum_{m = 1}^{\infty} 
\< 0 | gB_{i} (\vec{r}_{1}) | m\>
\<m | g B_{j}(\vec{r}_{2}) | 0\>
\nn\\*
&&\quad \times
e^{-(\Delta E_{m0})T/2} 
\cosh ((\Delta E_{m0})(T/2 -t))
\nonumber\\*
&&+O(e^{-(\Delta E_{10})T})\;.
\label{eqn:spectral-rep3}
\eea
After inserting these expressions into
Eqs.~\eqref{eqn:pot1}-\eqref{eqn:pot3-4},
we can carry out the integration and extrapolation, which
imply that
\bea
\lim_{\tau \to \infty} \int_{0}^\tau dt  \cdots
= \lim_{T \to \infty} \int_{0}^{\eta T} dt  \cdots 
\eea
with an arbitrary $\eta \in (0, 1/2]$.
Thereby we obtain the spectral representation of 
the spin-dependent potentials, which consists of
the matrix elements and the energy gaps.

\par
For the simplest parametrization
$\vec{r}_{1}=\vec{0}=(0,0,0)$ with $t_{1}=0$ and
$\vec{r}_{2}=\vec{r}=(r,0,0)$ with $t_{2}=t$, 
which is the actual setting of our simulation,
we have explicitly 
\bea
&&
V_{1}'(r)= 2  \sum_{m = 1}^{\infty}
\frac{\<0 | gB_{y}(\vec{0})|m  \> \<m  | gE_{z}(\vec{0})| 0\>}
{(\Delta E_{m0})^2}   \; ,
\label{eqn:pot1-final}\\
&&
V_{2}'(r)=2     \sum_{m = 1}^{\infty}
\frac{\<0 | gB_{y}(\vec{0})|m \> \<m | gE_{z}(\vec{r})| 0\>}
{(\Delta E_{m0})^2}   \; ,
\label{eqn:pot2-final}\\
&&  
V_{3}(r)=2 \sum_{m = 1}^{\infty} \! \left [
  \frac{\<0 | gB_{x}(\vec{0})|m \> \<m | gB_{x}(\vec{r})| 0\>}
{\Delta E_{m0}}   
- \frac{\<0 | gB_{y}(\vec{0})|m \> \<m | gB_{y}(\vec{r})| 0\>}
{\Delta E_{m0}}   \right ]
\; ,\label{eqn:pot3-final}\nn\\*
&&\\
&&
V_{4}(r)=2   \sum_{m = 1}^{\infty} \!\left [
\frac{\<0 | gB_{x}(\vec{0})|m \> \<m | gB_{x}(\vec{r})| 0\>}
{\Delta E_{m0}} 
+
2 \frac{\<0 | gB_{y}(\vec{0})|m \> \<m | gB_{y}(\vec{r})| 0\>}
{\Delta E_{m0}}    \right ] \; ,\nn\\*
\label{eqn:pot4-final}
\eea
where we have used the relations
\bea
&&\[ g^2 B_{y}(\vec{0}, 0) E_{z}(\vec{0}, t) \] 
=-\[ g^2 B_{z}(\vec{0}, 0) E_{y}(\vec{0}, t) \] \; , \\
&&\[ g^2 B_{y}(\vec{0}, 0) E_{z}(\vec{r}, t) \]
=-\[ g^2 B_{z}(\vec{0}, 0) E_{y}(\vec{r}, t) \] \; ,\\
&&\[ g^2 B_{y}(\vec{0}, 0) B_{y}(\vec{r},t) \]
=\[ g^2 B_{z}(\vec{0}, 0) B_{z}(\vec{r},t) \]\; .
\eea
We note that the error term in the field strength correlator of 
$O(e^{-(\Delta E_{10})T})$
in Eqs.~\eqref{eqn:spectral-rep1}-\eqref{eqn:spectral-rep3}
is assumed to be negligible, which is the case
for large~$T$.

\par
Now our procedure to compute 
the spin-dependent potentials is as follows;
we evaluate the field strength correlators
for various $r$ and $t$,
fit them to the spectral representation in
Eqs.~\eqref{eqn:spectral-rep1}-\eqref{eqn:spectral-rep3},
thereby determining the matrix elements and the energy gaps,
and insert them into 
Eqs.~\eqref{eqn:pot1-final}-\eqref{eqn:pot4-final}.
Here, the hyperbolic sine or cosine function in 
Eqs.~\eqref{eqn:spectral-rep1}-\eqref{eqn:spectral-rep3}
is typical for the PLCF, 
which automatically takes into account the effect 
of the finite temporal lattice size in the fit. 

\par
Note that if one uses the Wilson loop at this point,
the spectral representation is just a multi-exponential function and
the leading error term is of $O(e^{-(\Delta E_{10})(\Delta t)})$, where
$\Delta t$ is the relative temporal distance
between the spatial part of the Wilson loop
and the field strength operator~\cite{Bali:1996cj,Bali:1997am}.
Denoting the temporal extent of the Wilson loop by $T_{w}$,
one can fit the data in the range $t  \in [0,T_{w}-2 \Delta t]$, where
$T_{w}$ is at most $T/2$ because of the periodicity of the lattice volume.
Clearly the available fit range is more restricted
than in the PLCF case, even if
$\Delta t/a$ is chosen as small as possible,
say one or two.
It may be possible to suppress the
error term by applying smearing techniques to
the spatial links.
However, it is not immediately clear if this 
procedure really cures the error term.
At least, one needs fine tuning of the smearing
parameters and further systematic checks.

\subsection{Field strength operator on the lattice}
\label{subsect:field-strength}

On the lattice, we use the field strength operator defined
by $g a^2 F_{\mu\nu}(s) \equiv 
(U_{\mu\nu}(s)-U_{\mu\nu}^{\dagger}(s))/(2i)$,
where 
$U_{\mu\nu}(s)=
U_{\mu}(s)U_{\nu}(s+\hat{\mu})U_{\mu}^\dagger (s+\hat{\nu})
U_{\nu}^\dagger (s)$ is the plaquette variable at a 
site~$s=(s_{s},s_{t})$ with a spatial site $s_{s}$ and 
a temporal site  $s_{t}$.
We also define $U_{-\mu}(s)=U_{\mu}^\dagger(s-\hat{\mu})$.
Practically, we construct the color-electric and color-magnetic field 
operators by averaging the field strength operator as
\bea
&& ga^2E_{i}(s) = \frac{1}{2}ga^2\left(F_{4\, i}(s)\!+\!F_{-i \,4}(s)
\right) \; ,
\label{eqn:field-strength-e}\\
&& ga^2 B_{i}(s) =\frac{1 }{8} g a^2\epsilon_{ijk}\left(
F_{j\, k}(s)+F_{k \, -j}(s)+F_{-j\, -k}(s)+F_{-k \,j}(s)
\right) \; ,
\label{eqn:field-strength-b}
\quad\quad\quad
\eea
where we assume that $E_{i}(s)$ is defined on 
$(s_{s},s_{t}+1/2)$, and $B_{i}(s)$ on $(s_{s},s_{t})$, 
respectively.

\par
Now, as seen from Eqs.~\eqref{eqn:pot1-final}-\eqref{eqn:pot4-final},
the spin-dependent potentials consist not only of the energy gap
but also of the matrix element of the field strength operator,
and thus one needs to take into account the renormalization of the
latter.
This is due to the fact that the field strength operators depend
explicitly on the lattice cutoff $a$.
In the absence of a viable nonperturbative
renormalization prescription for the field 
strength operators in the presence of the quark-antiquark source,
we follow here the Huntley-Michael (HM)
procedure~\cite{Huntley:1986de}, which was also used
in Refs.~\cite{Bali:1996cj,Bali:1997am}.
This procedure is inspired by the weak coupling expansion 
of the Wilson loop and is aimed at removing the self-energy
contribution, at least, at $O(g^2)$.
We define $\bar{E}_{i}$ and $\bar{B}_{i}$ from $\bar{F}_{\mu\nu}(s) 
\equiv (U_{\mu\nu}(s)+U_{\mu\nu}^\dagger (s))/2$, and,
by taking the average according to Eqs.~\eqref{eqn:field-strength-e} 
and~\eqref{eqn:field-strength-b}, we compute
\bea
Z_{E_{i}} (r) = 1/\[ \bar{E}_{i} \]  \;,\quad
Z_{B_{i}} (r) = 1/ \[ \bar{B}_{i} \] \; ,
\label{eqn:HMfactor}
\eea
where $\bar{E}$ or $\bar{B}$ are attached to either side of 
the PLCF.
These factors are then multiplied to the field strength operators 
in Eqs.~\eqref{eqn:field-strength-e} and~\eqref{eqn:field-strength-b}
accordingly.
Note that $Z_{E_{i}}$ and $Z_{B_{i}}$ determined in this way 
can depend on $r$ and also on 
the relative orientation of the field strength operator to the 
quark-antiquark axis.

\par
One may find that the HM procedure is quite similar to tadpole 
improvement~\cite{Lepage:1992xa}, where the corresponding 
renormalization factor is defined by the inverse of the 
expectation value of the plaquette variables,
$Z_{\rm tad}=1/\<  U_{\Box}\>$, where
\bea
U_{\Box} = \frac{1}{6(L/a)^{3}(T/a)}\sum_{s,\mu >\nu}\frac{1}{3}
\mbox{Re~Tr}~U_{\mu\nu}(s) \; ,
\eea
which was used e.g. in Refs.~\cite{deForcrand:1985zc,Born:1993cp}.
Indeed, if the factorization of the correlator
$\< \bar{F}_{\mu\nu} \>_{PP^*} = \<\bar{F}_{\mu\nu} \>\< PP^* \>$ 
holds,\footnote{Numerically, this factorization is approximately satisfied.}
$Z_{B_{i}}$ and $Z_{E_{i}}$ are reduced to $Z_{\rm tad}$.
However, as was pointed out in \cite{Huntley:1986de}, the tadpole
factor does not remove the self energy even to $O(g^2)$ if
the correlator involves the electric field operator.

\subsection{Multi-level algorithm for the field strength
correlator}
\label{subsect:multi-level}

\par
We now describe the multi-level 
algorithm~\cite{Luscher:2001up,Luscher:2002qv} 
for computing 
the field strength correlators, restricting the discussion to the
lowest level. 
The essence of the multi-level algorithm is to 
construct the desired correlation function, which may have 
a very small expectation value, from the 
product of the relatively large ``sublattice average'' 
of its components.
We will also refer to such a component as the sublattice correlator.
\par
The sublattice is defined by dividing 
the lattice volume into several layers 
along the time direction, and
thus a sublattice consists of a certain number of time slices $\Ntsl$
(the number of sublattices is then $\Nsub=(T/a)/\Ntsl$, 
which is assumed to be an integer).
The sublattice correlators are evaluated
in each sublattice after updating the gauge field with a mixture
of heatbath~(HB) and over-relaxation~(OR) steps, 
while the space-like links on the boundary
between sublattices remain intact during the update.
We refer to this procedure as the ``internal update''.
We repeat the internal update $\Niupd$ times until we obtain a stable 
signal for the sublattice correlators.
Next, we multiply these  sublattice correlators  in a 
suitable way to complete the correlation function,
as described below.
Thereby the correlation function is obtained
for one configuration.
We then update the whole set of links without specifying any
layer to obtain another independent gauge configuration
and repeat the above sublattice averaging.
The computational cost for one configuration is
rather large, but one can observe a signal already from 
a few configurations once
$\Ntsl$ and $\Niupd$ are appropriately chosen.

\begin{table}[t]
\centering
\caption{A minimal set of 
sublattice correlators for the static potential and
the spin-dependent potentials.}
\vspace{0.3cm}
\begin{tabular}{lcl}
\hline
Potential &  & Sublattice correlators  \\
\hline
$V_{0}$ & &$\mathbb{T}_{PP}$\\
$V_{1}'$ &  & $\mathbb{T}_{PP}$, 
$\mathbb{T}_{PB_{y}}$, $\mathbb{T}_{PE_{z}}$, 
$\mathbb{T}_{P(B_{y}E_{z})}$\\
$V_{2}'$ &  & $\mathbb{T}_{PP}$,
$\mathbb{T}_{B_{y}P}$,   $\mathbb{T}_{P E_{z}}$, 
$\mathbb{T}_{B_{y}E_{z}}$\\
$V_{3}, V_{4}$ &  &$\mathbb{T}_{PP}$, 
$\mathbb{T}_{PB_{x}}$, $\mathbb{T}_{PB_{y}}$,
$\mathbb{T}_{B_{x}P}$, $\mathbb{T}_{B_{y}P}$,
$\mathbb{T}_{B_{x}B_{x}}$,
$\mathbb{T}_{B_{y}B_{y}}$\\
\hline
\end{tabular}
\label{tab:subcor}
\vspace*{0.5cm}
\end{table}

\par
In the current simulation,
the building blocks of the field strength correlators 
are the sublattice correlators listed in Table~\ref{tab:subcor}.
$\mathbb{T}$~represents the complex $9 \times 9$ matrices
that act on color tensors
in the ${\bf 3}\otimes {\bf 3}^*$ representation of SU(3).
The subscripts of $\mathbb{T}$ in Table~\ref{tab:subcor} 
denote the type of the sublattice correlators.
The way of completing a sublattice correlator is as follows.
Denoting the temporal sites as $s_{t}=(\itsl, \isub)$,
where $\itsl \in [1,\Ntsl]$ runs within the extent of one sublattice
labeled by $\isub \in [1,\Nsub]$,
a component of the Polyakov loop (timelike Wilson line
$\mathcal{P}$), the complex $3 \times 3$ matrices, 
in each sublattice is expressed as
\bea
\mathcal{P} (s_{s},\isub)_{\alpha\beta}
= \left ( \prod_{\jtsl =1}^{\Ntsl} 
U_{t}(s_{s}, \isub,\jtsl) \right )_{\alpha\beta}\; ,
\label{eqn:pline}
\eea
where we explicitly write the color labels in Greek letters.
The direct product of two timelike Wilson lines 
$\mathcal{P}$ separated by a distance $r$
is the simplest sublattice correlator, i.e.
\bea
\mathbb{T}_{P P}(s_{s}, \isub ;  r)_{\alpha\beta\gamma\delta}
=\mathcal{P}(s_{s},\isub)_{\alpha\beta} \otimes 
\mathcal{P}^{*} (s_{s}+r \hat{x},\isub)_{\gamma\delta}  \; ,
\eea
which is relevant to both the PLCF and the 
field strength correlators.

\par
Other sublattice correlators are constructed 
by inserting one or two field strength operators into the timelike 
Wilson line $\mathcal{P}$.
For instance, in order to obtain $\mathbb{T}_{PB_{y}}$,
$\mathbb{T}_{PE_{z}}$ etc., we compute in each sublattice 
the timelike Wilson line with a single 
insertion of the field strength
operator and then its direct product with $\mathcal{P}$.
The argument of this type of correlators is 
$(s_{s},\isub ; r, \itsl)$,
where $\itsl$ labels the timeslice where the field strength 
operator is inserted, $\itsl \in [1,\Ntsl]$. The quantities
$\mathbb{T}_{P(B_{y}E_{z})}$, 
$\mathbb{T}_{B_{y}E_{z}}$, 
$\mathbb{T}_{B_{x}B_{x}}$ and $\mathbb{T}_{B_{y}B_{y}}$
represent the double-field-strength-operator-inserted
sublattice correlators.
The argument of these correlators is
$(s_{s},\isub ; r,\ired)$, where 
$\ired \in [-\Ntsl +1 , \Ntsl -1]$
is the relative temporal distance between
two field strength operators.
For $\mathbb{T}_{P(B_{y}E_{z})}$, 
two field strength operators
are inserted into  one of two timelike Wilson lines, while 
for $\mathbb{T}_{B_{y}E_{z}}$,
$\mathbb{T}_{B_{x}B_{x}}$ and $\mathbb{T}_{B_{y}B_{y}}$,
they are inserted into both ones.

\par
The multiplication law of $\mathbb{T}_{PP}$ is then defined by
\bea
&&
\{ \mathbb{T}_{PP}(s_{s}, \isub ; r)
\mathbb{T}_{PP}(s_{s}, \isub +1 ; r)
\}_{\alpha\beta\gamma\delta}
\nn\\
&&=
\mathbb{T}_{PP}(s_{s}, \isub ; r)_{\alpha\rho\gamma\sigma}
\mathbb{T}_{PP}(s_{s}, \isub +1 ; r) _{\rho\beta\sigma\delta}
\label{eq:direct-product}
\eea
and this multiplication law of color components
is common to all other sublattice correlators.

\par
\begin{figure}[t]
\centering\includegraphics[width=8cm]{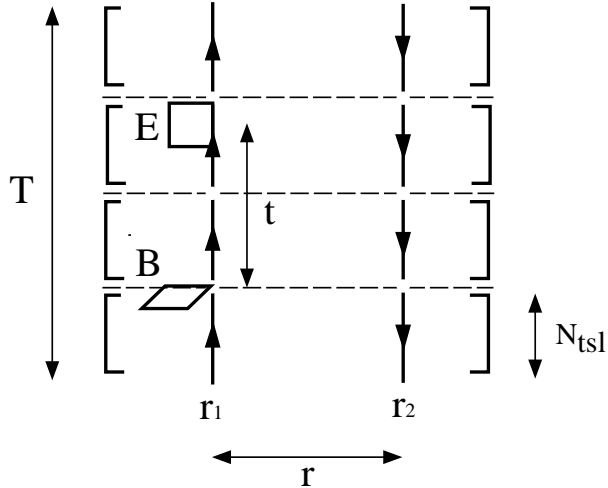}
\caption{How to complete 
$\[g^2 B_{y}(\vec{r}_{1},t_{1})E_{z}(\vec{r}_{1},t_{2})\]$
on the PLCF for $V_{1}'(r)$.
Arrows at $\vec{r}_{1}$ and $\vec{r}_{2}$ represent
the Polyakov lines for the static quark and antiquark.
$[\cdots]$ implies the sublattice average.
Other field strength correlators 
are obtained in a similar way.} 
\label{fig:fig1-be}
\end{figure}

After taking the sublattice averages,
we compute the PLCF for various $r$ by
\bea
P (\vec{0}) P^{*} (\vec{r})
&=& 
\frac{1}{9 (L/a)^3}\sum_{s_{s}}\mbox{Re}
\{ [\mathbb{T}_{PP}(s_{s}, 1 ; r) ]
[\mathbb{T}_{PP}(s_{s}, 2 ; r)] \times  \cdots
\nn\\
&&\qquad\times 
[\mathbb{T}_{PP}(s_{s}, \Nsub -1 ; r)]
[\mathbb{T}_{PP}(s_{s}, \Nsub ; r)]
\}_{\alpha\alpha\gamma\gamma} \; ,
\eea
and the field strength correlators 
for various $r$ and $t$
by combining other sublattice correlators,
where the translationally equivalent setting 
for space and time directions are averaged accordingly.
Figure~\ref{fig:fig1-be}
illustrates the computation of the field strength correlator
$\[ g^2 B_{y}(\vec{r}_{1},t_{1})E_{z}(\vec{r}_{1},t_{2})\]$
for $V_{1}'(r)$ (see also 
ref.~\cite{Koma:2003gi} for a similar
application of the multi-level algorithm,
in which the electric-flux
profile between static charges was measured with the PLCF).

\par
In order to benefit from the multi-level algorithm,
we need to optimize $\Ntsl$ and $\Niupd$.
They depend on the coupling $\beta$ and on
the distances to be investigated, which can be determined by 
looking at the behavior of the correlation function as a 
function of~$\Niupd$ for several~$\Ntsl$.
As an empirical observation we note that 
$a \Ntsl = 0.3 - 0.4$~fm is optimal in order 
to suppress the statistical fluctuation 
of the correlation functions.

\par
In principle, one can apply the
above computation to any direction of the 
quark-antiquark axis,
$\vec{r}=(r_{x},r_{y},r_{z})$ with 
$r=\sqrt{r_{x}^2 +r_{y}^2+r_{z}^2}$.
However, one needs to take into account the fact that 
even with the simplest parametrization, $\vec{r}=(r,0,0)$,
this algorithm requires a lot of memory 
to keep all $\mathbb{T}$ listed in Table~\ref{tab:subcor}
during the internal update.
For a fixed distance and a fixed quark-antiquark axis
using single precision,
$\mathbb{T}_{PP}$ requires memory space 
$(L/a)^3 \times \Nsub \times 162 \times 4$~bytes 
($\equiv 1$ work space unit {\it wsu}).
Therefore, to compute all spin-dependent potentials 
in the same setting, one needs additionally
$5 \Ntsl$~{\it wsu} for the single- and 
$4 (2 \Ntsl -1)$~{\it wsu} for the double-field-strength-inserted 
sublattice correlators.
For instance, on the $20^{3} 40$ lattice with 
$\Ntsl = 4$ ($\Nsub =8$),
about $49~{\it wsu}~=2$~GB memory
is needed as $1~{\it wsu} = 42$~MB.

\par
The way of obtaining the HM factors $Z_{E_{z}}$, $Z_{B_{y}}$
and $Z_{B_{z}}$ using the multi-level algorithm
is the same as above; we evaluate the sublattice average of
$\mathbb{T}_{P\bar{B}_{x}}$, $\mathbb{T}_{P\bar{B}_{y}}$
and $\mathbb{T}_{P\bar{E}_{z}}$ and complete the correlators
in Eq.~\eqref{eqn:HMfactor}.
The additional memory requirement is $3 \Ntsl$~{\it wsu} 
in the above setting.

\section{Numerical results}
\label{sec:result}

\par
In this section, we present our numerical results.
Simulation parameters are summarized in 
Table~\ref{tbl:simulation}.
We investigated the bare gauge couplings
$\beta=6.0$ ($a \approx 0.093$ fm) on 
several lattice volumes
and $\beta=6.3$ ($a \approx 0.059$ fm) 
on a~$24^4$ lattice.
The physical lattice spacing $a$ was fixed in terms of the Sommer
scale by setting $r_{0}=0.5$ fm~\cite{Necco:2001xg}. 
One Monte Carlo update consisted of a combination of 
1~HB  and 5~OR steps.
The number of internal updates $\Niupd$ for each $\beta$ value 
was optimized empirically to obtain signals up to intermediate 
distances.
We note that $\Ntsl$ and $\Niupd$ could further be tuned so that 
even larger distances become accessible.
The statistical errors were estimated
by applying the single elimination jackknife analysis.
The various fit parameters were determined by minimizing
$\chi^{2}$ defined with the full covariance matrix,
and their errors were estimated from the 
distribution of the jackknife samples.
For a consistency check we also evaluated the errors of the
fit parameters from the minimum value of the 
$\chi^{2}$ through $\Delta\chi_{\rm min}^{2}=1$.
In general, the errors from the jackknife analysis
were the same or slightly larger 
compared to those
estimated from $\Delta\chi_{\rm min}^{2}=1$.

\begin{table}[t]
  \caption{Simulation parameters used in this study.
The fourth column denotes the available quark-antiquark
distances for the static potential $V_{0}$ (before tree-level improvement),
while $[\cdots ]$ applies to the spin-dependent potentials
$V_{1}'$, $V_{2}'$, $V_{3}$ and $V_{4}$.}
\vspace{0.3cm}
\centering
\begin{tabular}{cccccrc}
\hline
$\beta=6/g^2$ 
& $a$~[fm] & $(L/a)^3 (T/a)$  & $ r/a$ & $N_{\rm tsl}$ &
$N_{\rm iupd}$ & $N_{\rm conf}$ \\
\hline
6.0 &  0.093 & $16^{4}\F\F$&  $2-7~[2-6]$  & 4 & 10000 & 90 \\
 6.0    &  0.093 & $20^{4}\F\F$&  $2-9~[2-6]$  & 4 &  7000 & 82 \\
 6.0    & 0.093  & $20^{3}40$&  $2-9~[2-7]$& 4 &  7000 & 33\\
6.3  & 0.059  & $24^{4}\F\F$& $2-8~[2-6]$ & 6 & 6000 & 39 \\
\hline
\end{tabular}
\label{tbl:simulation}
  \vspace*{0.3cm}
\end{table}

\subsection{Static potential and its derivatives}

\begin{figure}[t]
\centering\includegraphics[width=9.5cm]{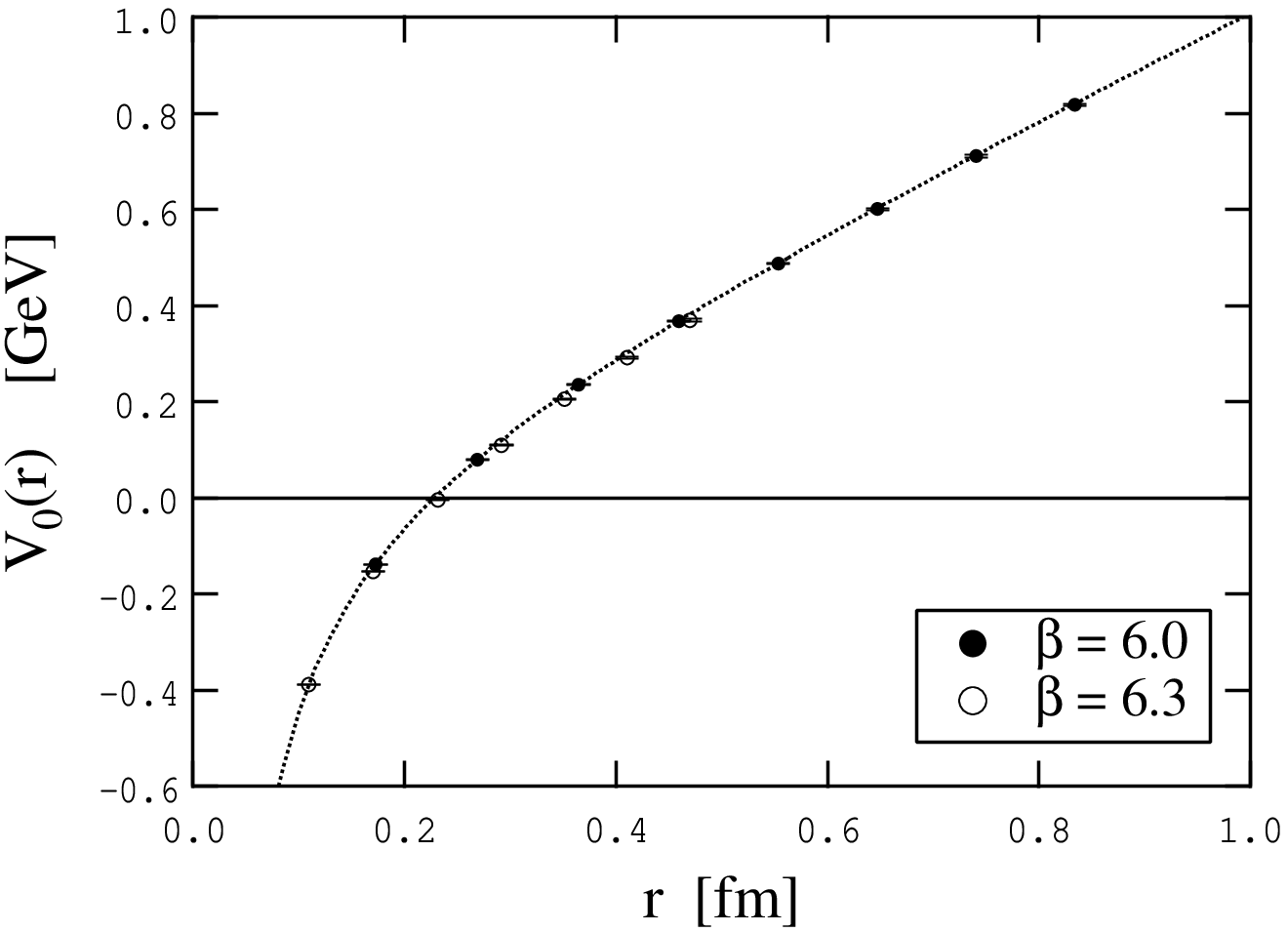}
\caption{Static potential $V_{0}(r_{I})$
at $\beta=6.0$ on the $20^{3}40$ lattice
and at $\beta=6.3$ on the $24^{4}$ lattice.
The constant term is subtracted.
The dotted line is the fit curve
Eq.~\eqref{eqn:potential-fit},
applied to the data at $\beta=6.0$.} 
\label{fig:static_potential}
\centering\includegraphics[width=9.5cm]{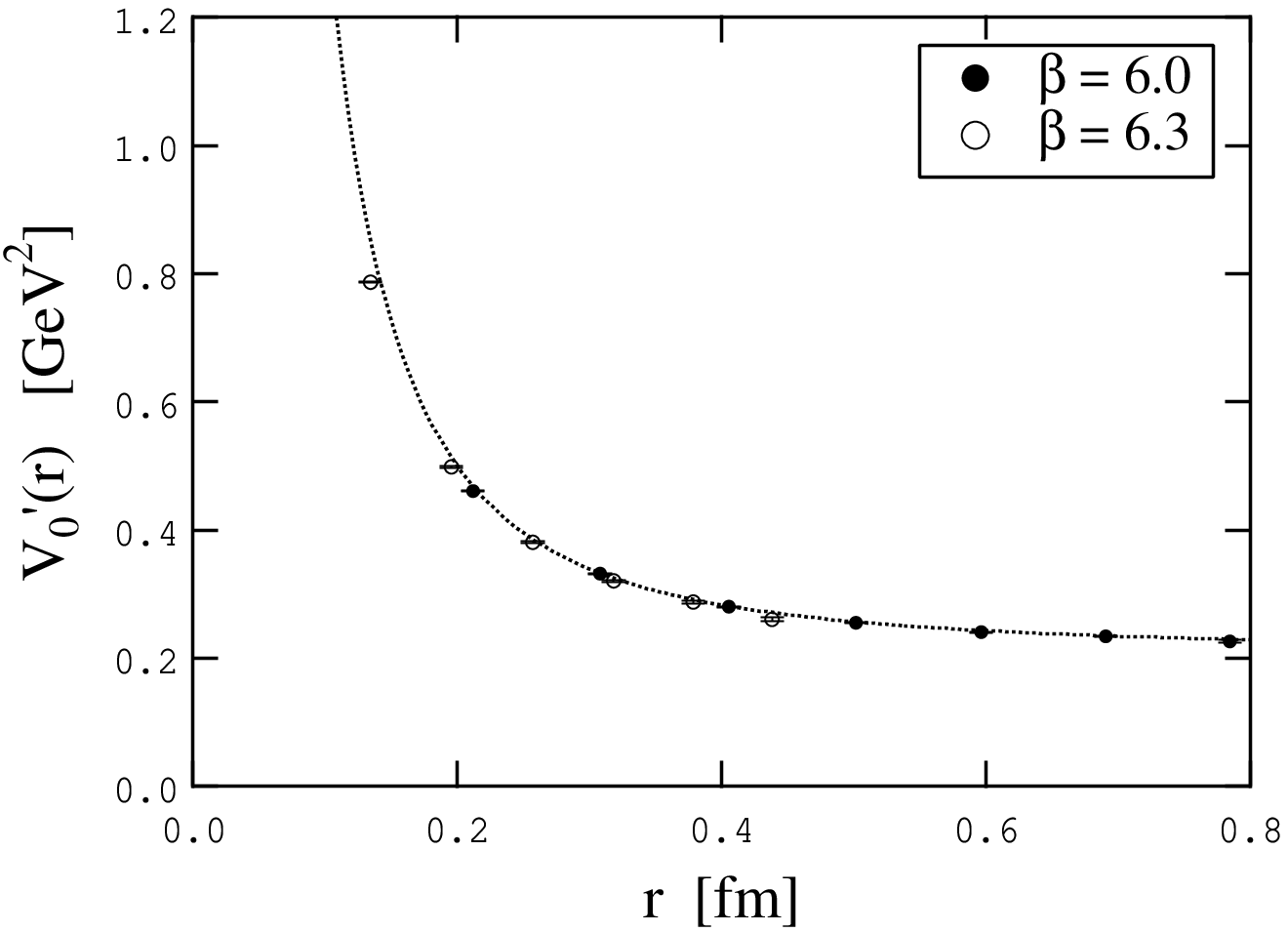}
\caption{The force $V_{0}'(\bar{r})$.
The dotted line is the fit curve
Eq.~\eqref{eqn:force-fit},
applied to the data at $\beta=6.0$.} 
\label{fig:force}
\vspace*{0.3cm}
\end{figure}

\par
\begin{figure}[t]
\centering\includegraphics[width=9.5cm]{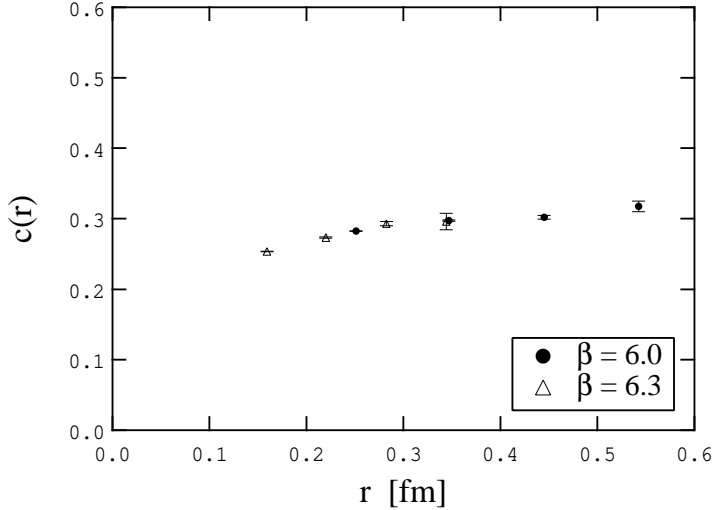}
\caption{$c(\tilde{r})= - \tilde{r}^3 V_{0}''(\tilde{r}) /2$.} 
\label{fig:cr}
\vspace*{0.3cm}
\end{figure}

We first present the basic quantities extracted from the
PLCF, i.e. the static potential and its derivatives
with respect to the distance, 
in Figs.~\ref{fig:static_potential}-\ref{fig:cr}, 
which are defined by
\bea
V_{0}(r_{I})
&=& -\frac{1}{T} 
\ln \<P(\vec{0})P^{*}(\vec{r})\> + O(e^{-(\Delta E_{10})T}) \;, 
\\*
V_{0}'(\bar{r})
&=& \frac{1}{a}\left \{ V_{0}(r)-V_{0}(r-a)\right \}\; ,
\label{eqn:forcedef} \\*
 \frac{1}{2}\tilde{r}^{3} V_{0}''(\tilde{r})
&=&
\frac{1}{2}\tilde{r}^{3} 
\frac{1}{a^{2}}\left 
\{ V_{0}(r+a)+V_{0}(r-a)-2V_{0}(r)\right \}
\equiv  -c(\tilde{r}) \; .
\eea
We have applied tree-level improvement to
the quark-antiquark distances in order to avoid
an enhancement of lattice discretization
effects especially at small 
distances~\cite{Sommer:1993ce,Necco:2001xg,Luscher:2002qv},
so that the distances, $r_{I}$, $\bar{r}$ and $\tilde{r}$ 
are defined through the relations
\bea
&&
r_{I}^{-1}=4\pi G(r)\; ,\\
&&
\bar{r}^{-2}= \frac{4\pi}{a}\left\{G(r-a)-G(r)\right \}\; ,
\\
&&\tilde{r}^{-3}=  \frac{2\pi}{a^{2}}
\left\{ G(r+a)+G(r-a)-2G(r)\right\}\;,
\eea
where $G(r)\equiv G(r,0,0)$ is
the Green function of the lattice Laplacian in three
dimensions.
For convenience we summarize these distances 
in Table~\ref{tbl:distance} in the Appendix.

\par
We then fit the potential 
and the force (first derivative of the potential) to the functions
\bea
&&
V_{\mathrm{fit}}(r) = \sigma r - \frac{c}{r} + \mu \; ,
\label{eqn:potential-fit}\\
&&
V'_{\mathrm{fit}}(r) = \sigma  + \frac{c}{r^{2}} \; ,
\label{eqn:force-fit}
\eea
and estimate the string tension $\sigma$,
the Coulombic coefficient $c$, and a constant~$\mu$.
The fit results at $\beta=6.0$ are summarized in Table~\ref{tbl:potfit}.
We find that the string tensions determined from either the 
potential or the force are consistent with each other.
There is a small difference in $c$. 
However, this may be acceptable, 
since we see that $c(\tilde{r})$, which is extracted from the
second derivative of the static potential,
is not strictly constant as a function of $r$ as shown 
in Fig.~\ref{fig:cr}.
Thus $c$ can be affected by the additional fit terms.
Also, given that the estimates for $c$ have not
stabilized in the considered
range of distances, it is not too surprising that a fit
ansatz based on Eqs.~\eqref{eqn:potential-fit} 
and \eqref{eqn:force-fit} produces a large value of $\chi^2$.
Nevertheless, it is interesting to find 
that the fit curves characterize the global feature of 
the potential and the force.
In these fits, we found no strong dependence on 
the fit range.
We also examined the force obtained by 
the central derivative $V_{0}'(\bar{r}_{c})=
\{ V_{0}(r+a)-V_{0}(r-a)\} /(2a)$, where $\bar{r}_{c}$
is defined via
$\bar{r}_{c}^{-2}= 4 \pi \{ G(r-a)-G(r+a) \}/ (2a)$,
and found the same curve as in Fig.~\ref{fig:force}.
Later, we shall compare the values of $\sigma$ and $c$ with 
those extracted from the spin-dependent potentials.

\begin{table}[t]
\centering
\caption{Fit results of the static potential and the force
at $\beta=6.0$ on the $20^3 40$ lattice with 
the fit functions in 
Eqs.~\eqref{eqn:potential-fit} and \eqref{eqn:force-fit}.
The corresponding fit curves are plotted in 
Figs.~\ref{fig:static_potential} and~\ref{fig:force}.}
\vspace*{0.3cm}
\begin{tabular}{cccccc}
\hline
& Fit range ($r/a$) &   $\sigma a^2$ & $c$  & $\mu a$ &
$ \chi_{\rm min}^{2}/N_{\rm df}$ 	\\
\hline
$V_{0}(r)$ &
$1.855-8.971$ &   0.0466(2)  &0.281(5) &  0.7169(5) &   3.8 \\
$V_{0}'(r)$
&$3.312-8.438$&   0.0468(2) & 0.297(1) & --- & 5.6\\
\hline
\end{tabular}
\label{tbl:potfit}
\vspace*{0.25cm}
\end{table}

\par
At large enough distances, one may expect the value 
$c= \pi/12 \approx 0.262$ from the bosonic string theory
in four dimensions~\cite{Luscher:1980fr,Luscher:1980ac}. 
However, as is clear from the plot of $c(\tilde{r})$ in Fig.~\ref{fig:cr}, 
we find $c(\tilde{r}) \approx 0.3$ at 
$r\gtrsim 0.3$~fm, which differs from this expectation 
by about 13~\%.
To accommodate a value of $\pi/12$ for the coefficient of the $1/r$ term,
one would need higher order corrections at these distances.
Note that the results for  $c(\tilde{r})$ obtained here 
are consistent with Refs.~\cite{Luscher:2002qv,HariDass:2006pq}.

\subsection{HM renormalization factors}

\begin{figure}[t]
\centering\includegraphics[width=7cm]{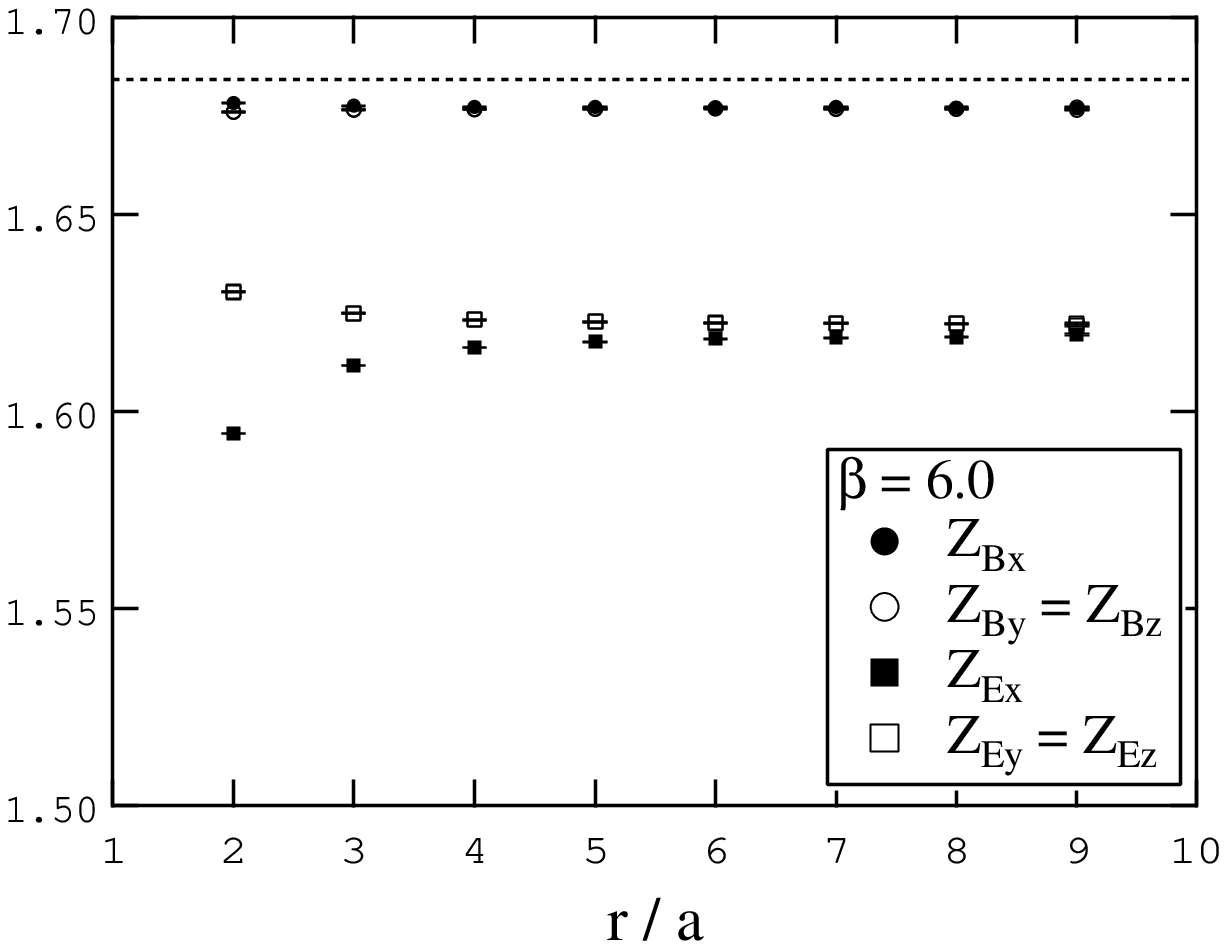}%
\includegraphics[width=7cm]{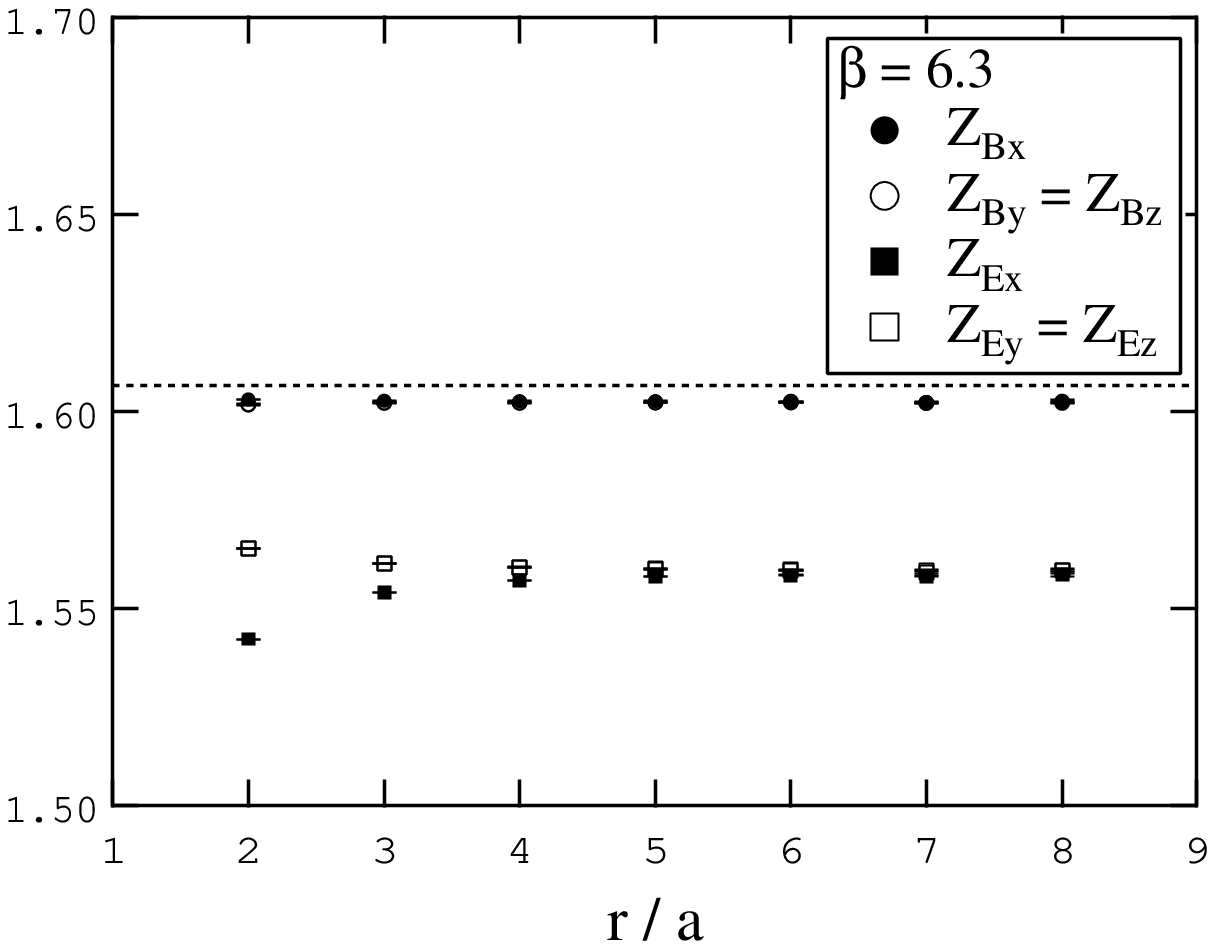}
\caption{The HM  factors 
at $\beta=6.0$ (left) and $\beta=6.3$ (right)
as a function of~$r$.
The dashed lines correspond to the tadpole estimate from 
the inverse of the expectation value of the plaquette:
$Z_{\rm tad}=1.684$ ($\< U_{\Box}\> 
=0.59373(4)$) at $\beta=6.0$
and $Z_{\rm tad}=1.607$ ($\< U_{\Box}\> 
=0.62241(2)$) at $\beta=6.3$.} 
\label{fig:HMfactor}
\end{figure}

In Fig.~\ref{fig:HMfactor}, we show 
the HM renormalization factors defined in Eq.~\eqref{eqn:HMfactor},
together with the tadpole renormalization 
factor (see Table~\ref{tbl:HMfactor} 
in the Appendix for the numerical values).
$Z_{B_{i}}$ are almost constant as a function of $r$, while
$Z_{E_{i}}$ exhibit some dependences on $r$ and on 
the relative orientation 
of the operator to the quark-antiquark axis at smaller distances.
The HM factors are generally smaller than the tadpole
factor by less than 1~\% for $Z_{B_{i}}$ 
and at most 6~\% for $Z_{E_{i}}$.
Since the statistical fluctuations of these factors are 
much smaller than that of the field strength correlators,
we ignore their errors when we multiply them
to the field strength correlators, but 
take into account their $r$-dependences.
Note that the observed tendencies are in agreement 
with Ref.~\cite{Bali:1997am}, where the Wilson loop was used.

\subsection{Field strength correlators}

\begin{figure}[tb]
\centering
\includegraphics[width=7cm]{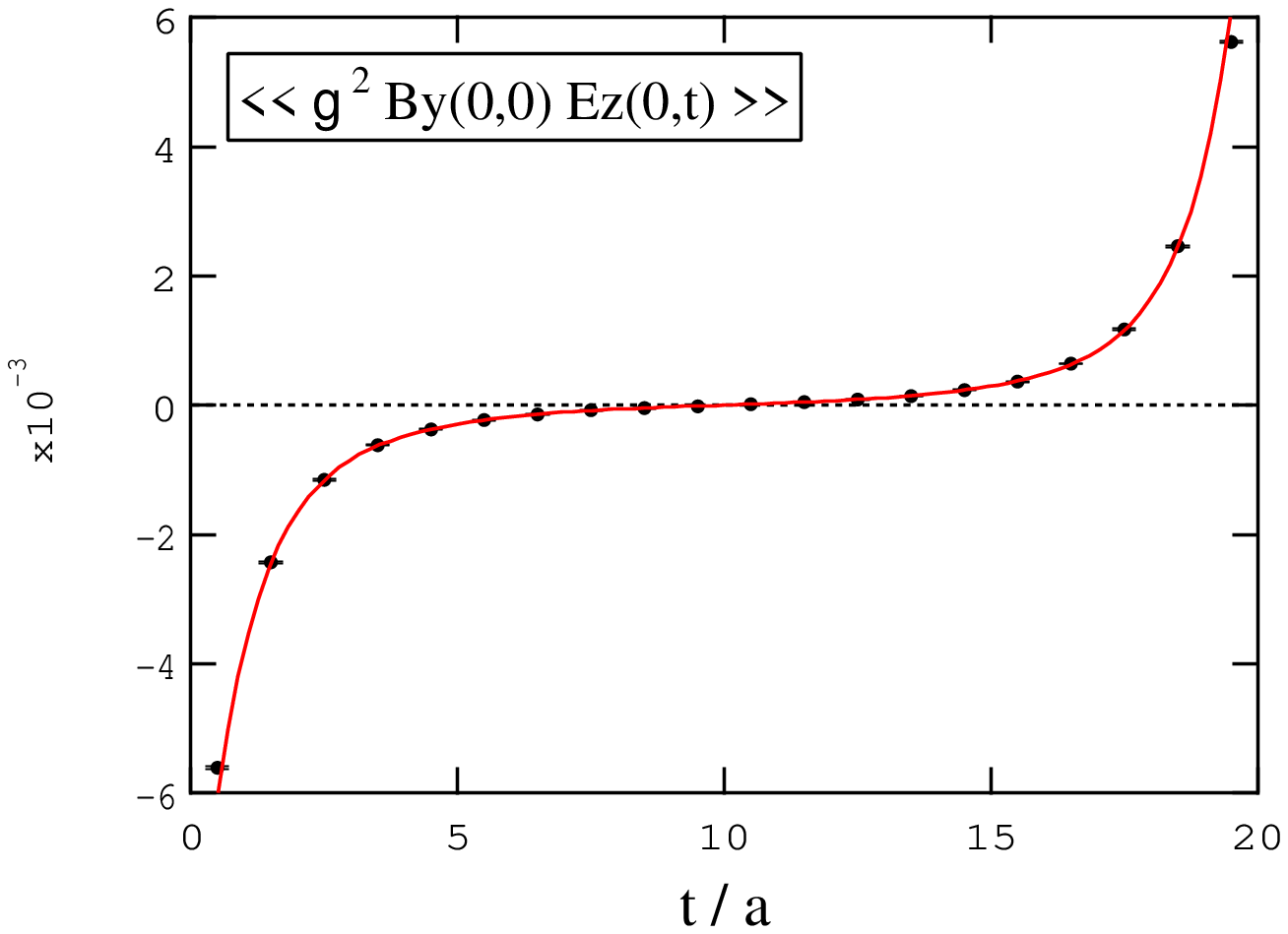}%
\includegraphics[width=7cm]{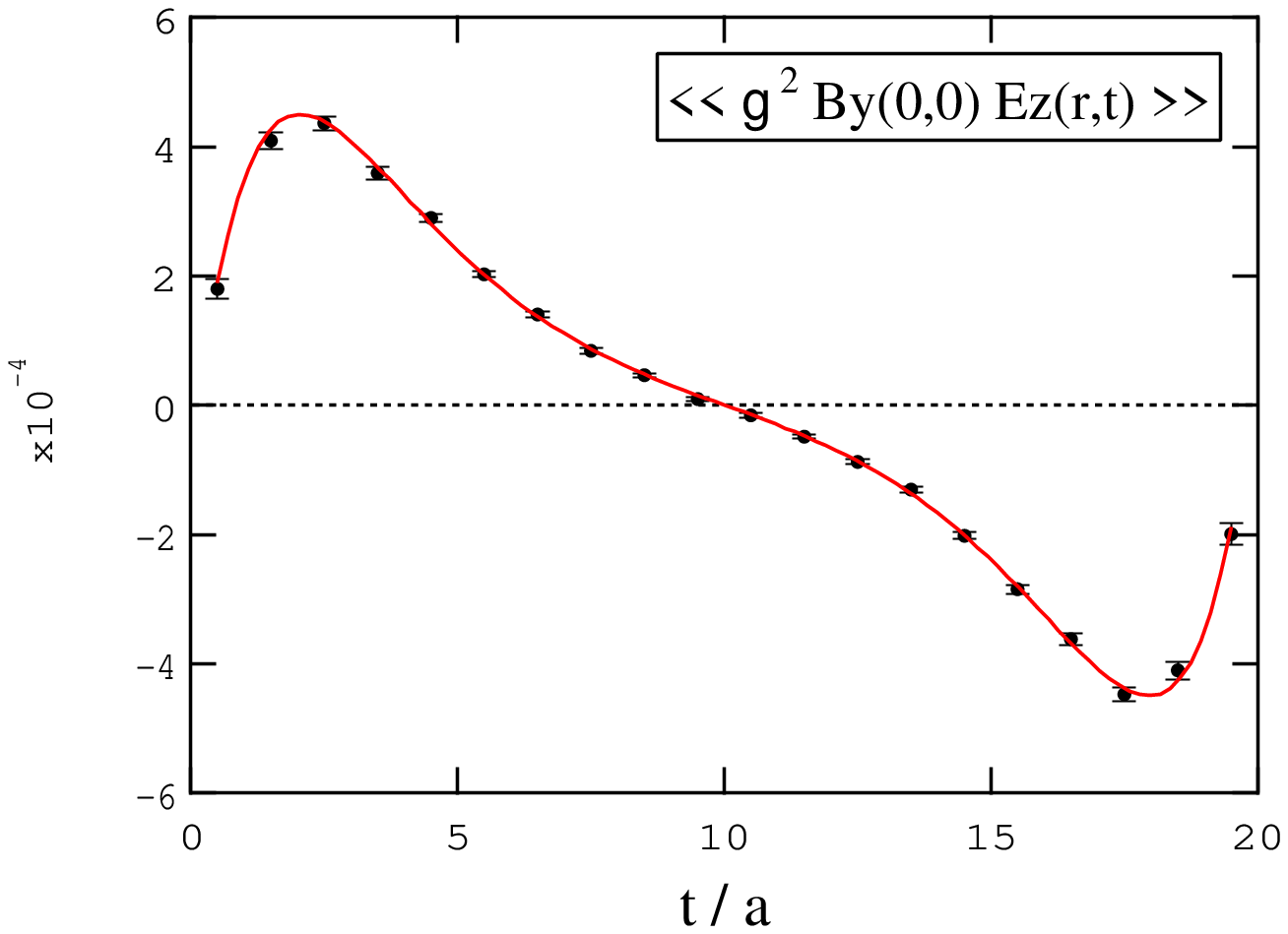}
\centering
\includegraphics[width=7cm]{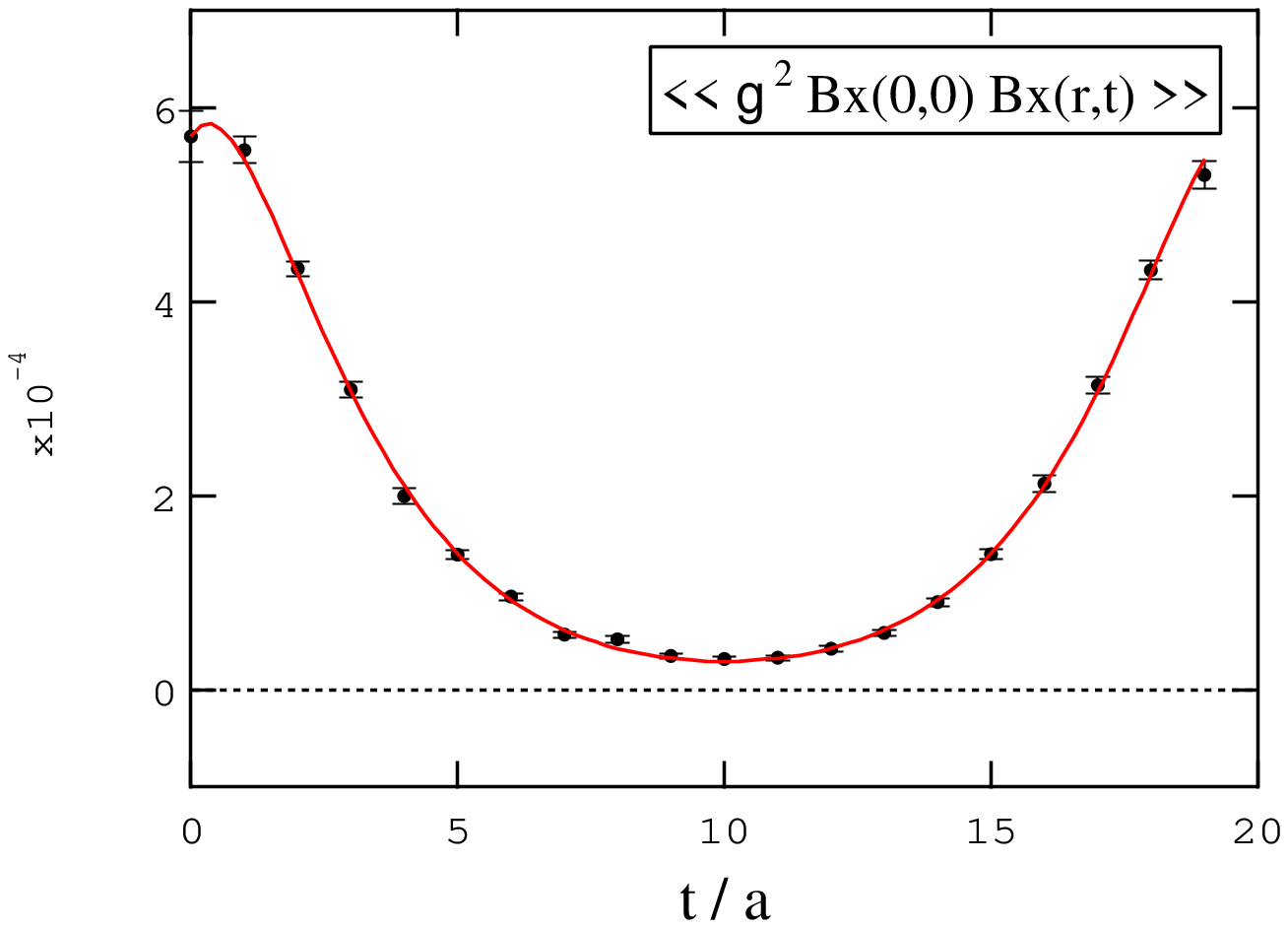}%
\includegraphics[width=7cm]{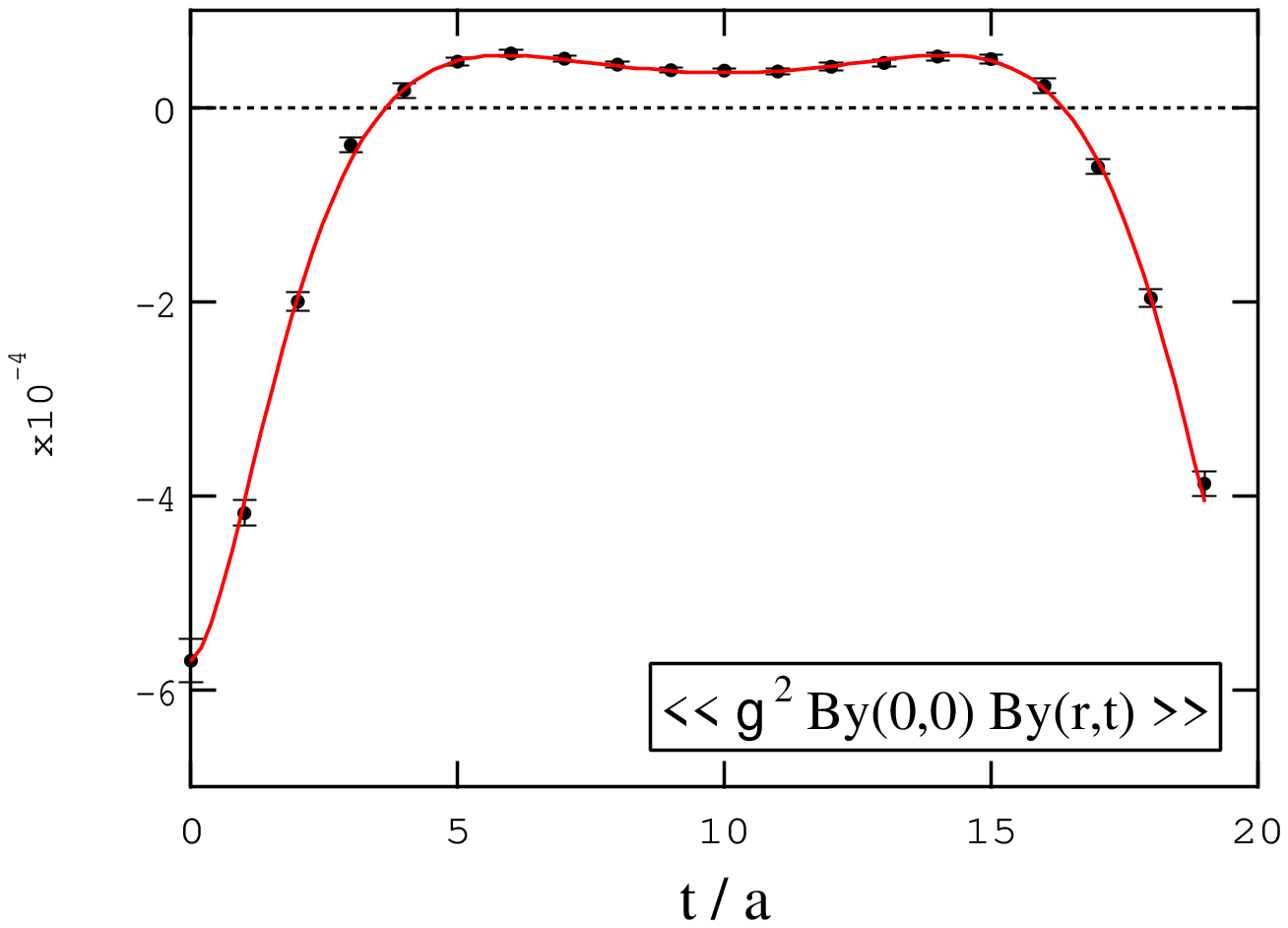}
\caption{Field strength correlators at $\beta=6.0$ on the
$20^4$ lattice for $r/a=5$ as a function of $t/a$.
The solid lines are the fit curves corresponding to
Eqs.~\eqref{eqn:spectral-rep1}-\eqref{eqn:spectral-rep3}.}
\label{fig:correlators20}
\vspace*{0.5cm}
\end{figure}

\begin{figure}[tb]
\centering
\includegraphics[width=7cm]{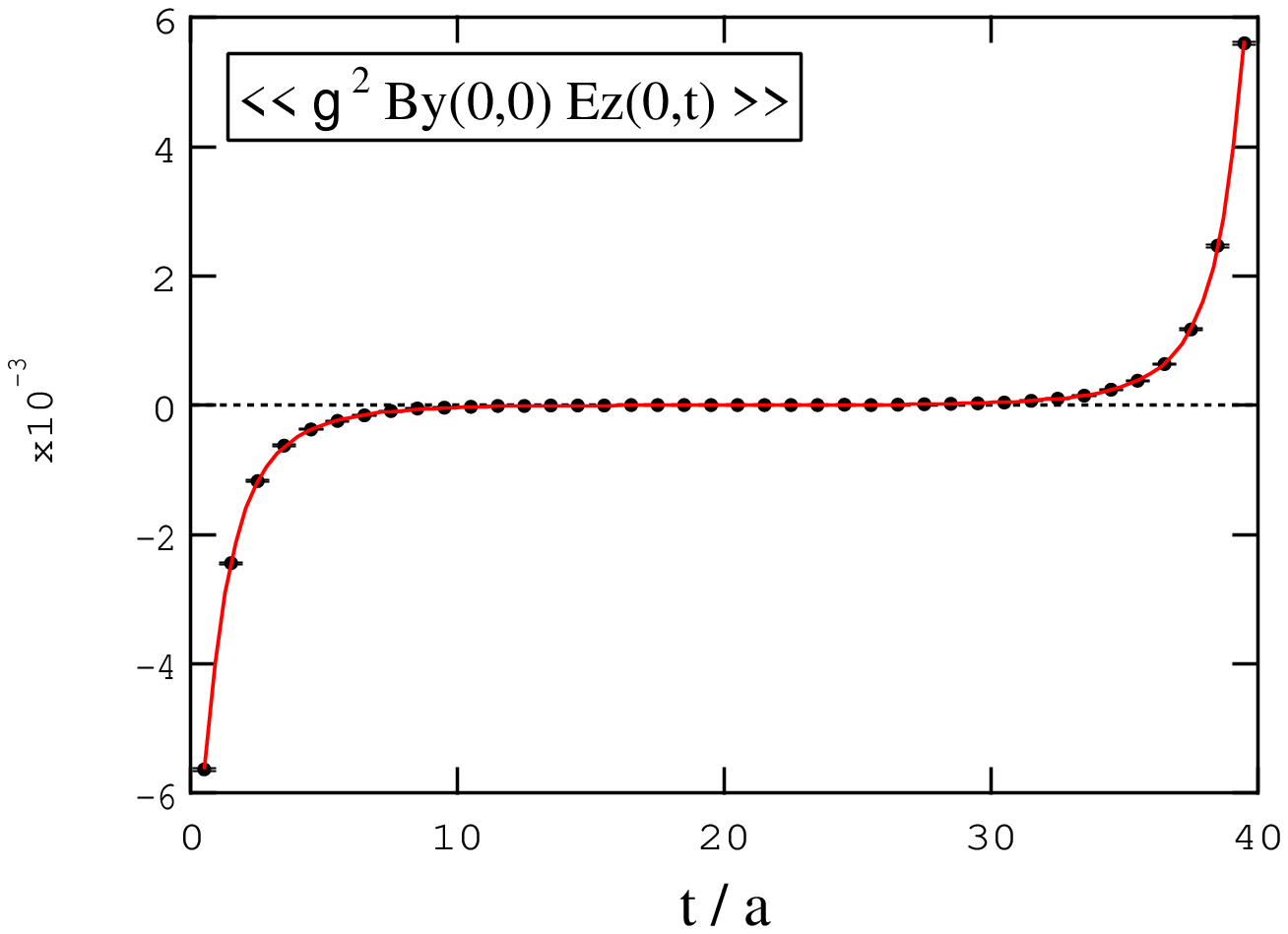}%
\includegraphics[width=7cm]{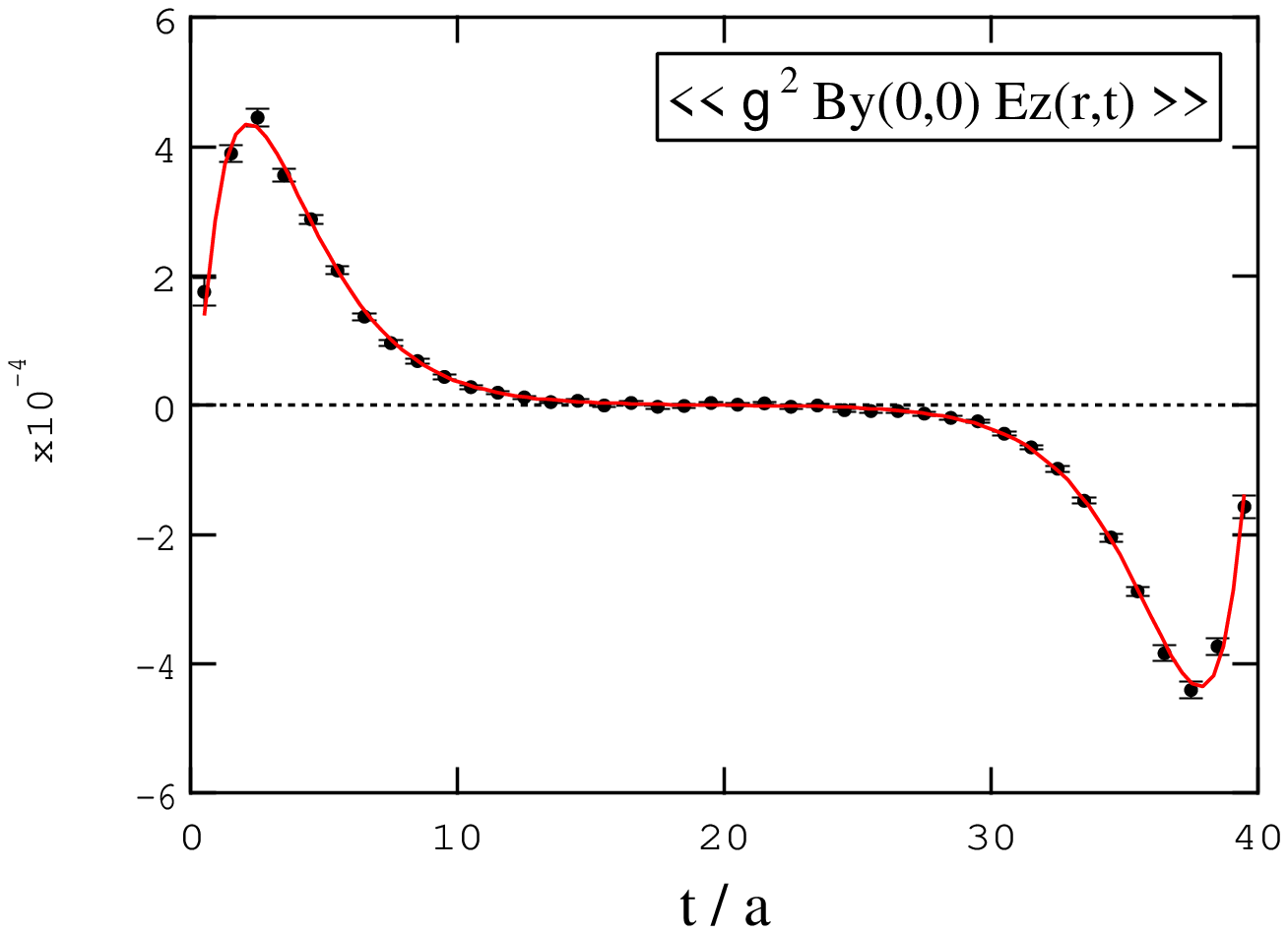}
\centering
\includegraphics[width=7cm]{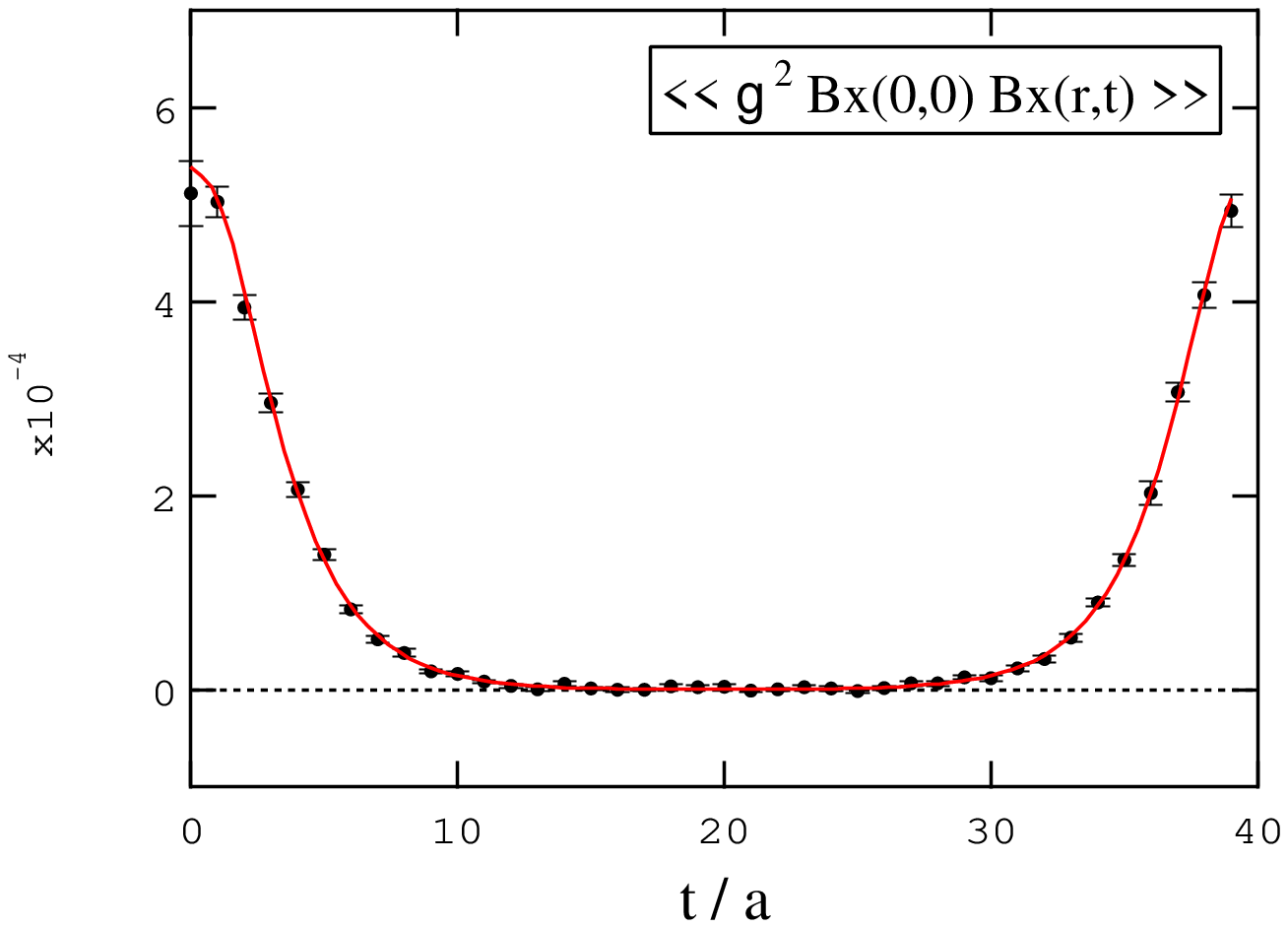}%
\includegraphics[width=7cm]{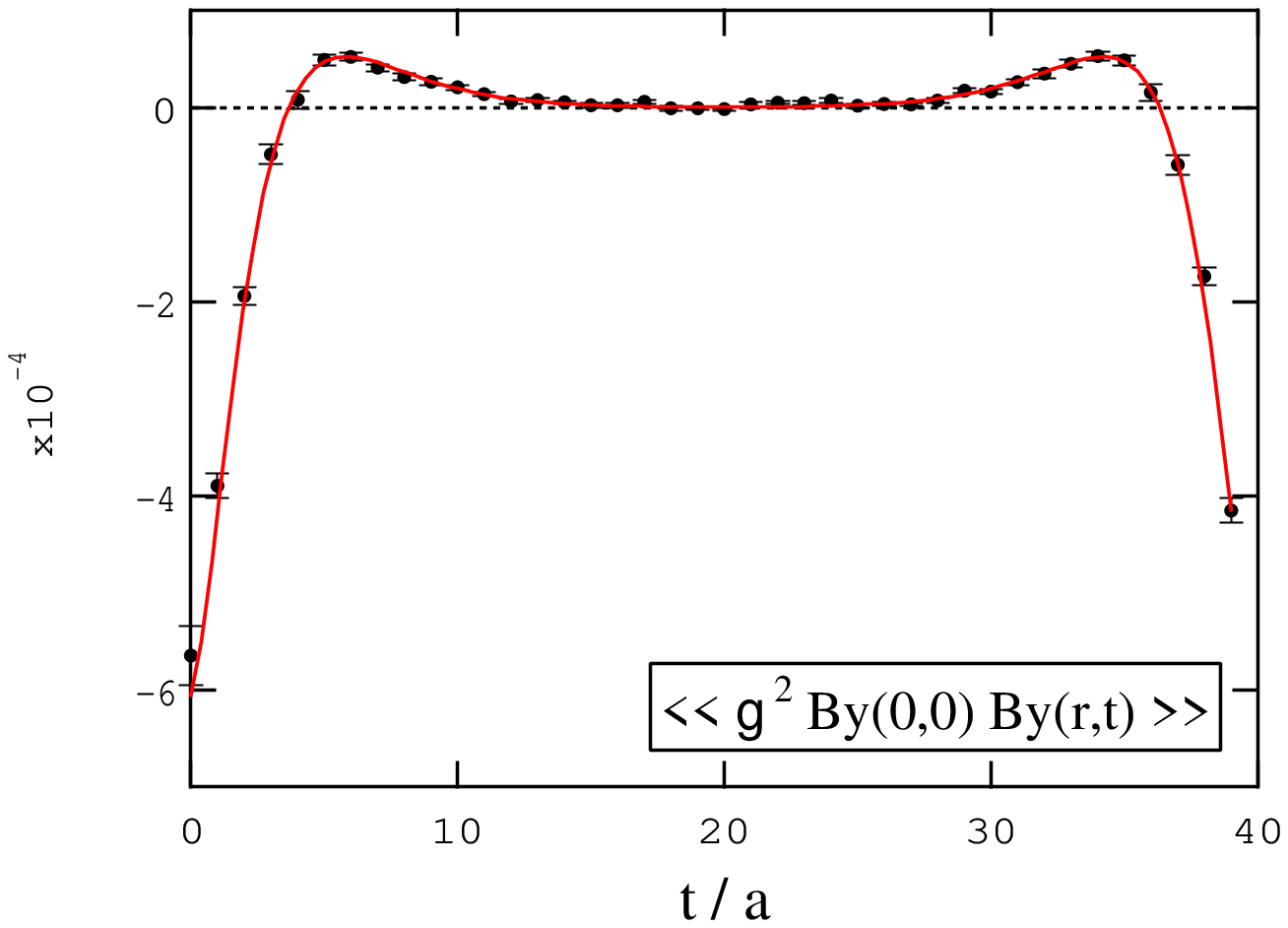}
\caption{The same as Fig.~\ref{fig:correlators20} but
on the $20^3 40$ lattice.}
\label{fig:correlators2040}
\vspace*{0.5cm}
\end{figure}

In Figs.~\ref{fig:correlators20} and~\ref{fig:correlators2040}, 
we show the various field strength correlators
as a function of $t$ at $\beta=6.0$ on
the $20^4$ and $20^{3}40$ lattices, respectively, where
$r/a=5$ is selected as an example.
At smaller distances, the quality of the data is even better.
Owing to the multi-level algorithm, the statistical accuracy of the 
data is unprecedented,
which allows us see the typical behavior of correlators.

\par
We also include the fit curves in these figures, which are
supplied by
the spectral representation of the field strength 
correlators on the PLCF 
in subsection~\ref{subsect:spectral-rep}.
We find that 
Eqs.~\eqref{eqn:spectral-rep1}-\eqref{eqn:spectral-rep3}
provide an excellent description of
the behavior of the lattice data.
Our fit procedure was as follows.
We first folded the data
of $C(t)=\[g^2 B_{y}(\vec{0},0)E_{z}(\vec{0},t)\]$
and $\[g^2 B_{y}(\vec{0},0)E_{z}(\vec{r},t)\]$
as $\{ C(t)-C(T-t)\}/2 \to C(t)$ with
$t \in [0.5a,(T-a)/2]$,
and the data
of $C(t)=\[g^2 B_{x}(\vec{0},0)B_{x}(\vec{r},t)\]$
and $\[g^2 B_{y}(\vec{0},0)B_{y}(\vec{r},t)\]$
as $\{C(t) + C(T-t)\}/ 2 \to C(t)$
with $t \in [0,T/2]$.
Then, we fitted all available data points 
for each correlator in order to take into account
as many excited states as possible
in the spectral representations, except for $t/a=0.5$ in 
$\[ g^2 B_{y}(\vec{0}, 0) E_{z}(\vec{0},t) \]$, which was
to avoid unwanted lattice effects due to the
sharing of the same link variable in the two
field strength operators.
Here, since it is impossible to fix the parameters,
matrix elements and the energy gaps,
for all excited states, $m \geq 1$, with a finite number
of data points, 
we truncated the expansion at a certain $m=m_{\rm max}$.
The validity of this truncation was verified by
monitoring the values of $\chi^2$ and the integration results.
We generally chose $m_{\rm max}$  
such that $\chi_{\rm min}^2/N_{\rm df}$ was of order~1, where
$N_{\rm df}$ is the number of degrees of freedom.
All fit details and the fit results
(i.e. the values of the integral)
are tabulated in 
Tables~\ref{tbl:corfit-L20b60} and~\ref{tbl:corfit-L20T40b60}  
in the Appendix.

\par
An important observation is that the integrals obtained for
$T=20a$ and $T=40a$ are the same within errors,
despite the fact that the behavior of the correlators
around $t=T/2$ in Figs.~\ref{fig:correlators20}
and~\ref{fig:correlators2040} is obviously different.
While the correlator computed for $T=20a$ is clearly distorted
due to periodicity, this is not the case for $T=40a$
(note that the vertical axis is the same in both figures).
This indicates that the hyperbolic sine or cosine
function in the spectral representation provides
a good description of the finite-$T$ effect.
In other words, it is possible to
extract the asymptotic value of the integral from 
the lattice with a relatively small value of $T$ within this approach.
At the same time, this confirms that
the error term of $O(e^{-(\Delta E_{10})T})$ is 
negligible in this setting.
We also examined a smaller lattice volume, $16^{4}$,
at $\beta=6.0$ and obtained the same result
at $r/a \le 6$.\footnote{These data are not presented in this paper
but available on request.}
In this sense the spatial finite volume effect at $r/a \le 6$
on the $20^4$ and $20^3 40$ lattices is also negligible.

\par
The fit result at $\beta=6.3$ on the $24^4$ lattice 
is  listed in Table~\ref{tbl:corfit-L24b63} 
in the Appendix.
The finite volume effect at this $\beta$ value
is expected to be small, though 
we did not investigate this explicitly,
since the physical size of the lattice volume is 
almost the same as for~$16^{4}$ at~$\beta=6.0$.

\par
Finally, we point out several caveats.
i) Although the fit works beautifully, one may not be able 
to assign a quantitative meaning to the resulting matrix
elements and the energy gaps.
This is because we truncated the spectral representation
when performing the fit, and as a result, the lattice data, 
which in principle contains the contribution from
all modes, are forced to be described by only a few modes.
In this case, it is reasonable to regard
only the value for the integral to be of quantitative
significance. 
ii) As the energy gaps of the various field strength correlators in 
Eqs.~\eqref{eqn:spectral-rep1}-\eqref{eqn:spectral-rep3}
are the same, one may expect that 
the fit of each correlator provides
the same energy gap,
or one may attempt a simultaneous fit of
all correlators with such a constraint.
However, as the effective truncation level
is not always common to all correlators,
even at a fixed distance, which is 
also related to the fact that the matrix 
elements are not positive definite,
this was not always the case.
iii) The sensitivity of the fit result, namely the 
integration value, is mostly governed by the 
lowest energy gap selected by the fit, which gives
the dominant contribution at $\tau \to \infty$.
This is why we examined two lattice volumes with
the same spatial size but different temporal extent
and confirmed that such a systematic effect
is negligible.
This fact supports our claim that the spectral representation 
of the field strength correlator is useful
even though a truncation must necessarily
be performed.

\subsection{Spin-dependent potentials}

\par
\begin{figure}[t]
\centering
\includegraphics[width=9.5cm]{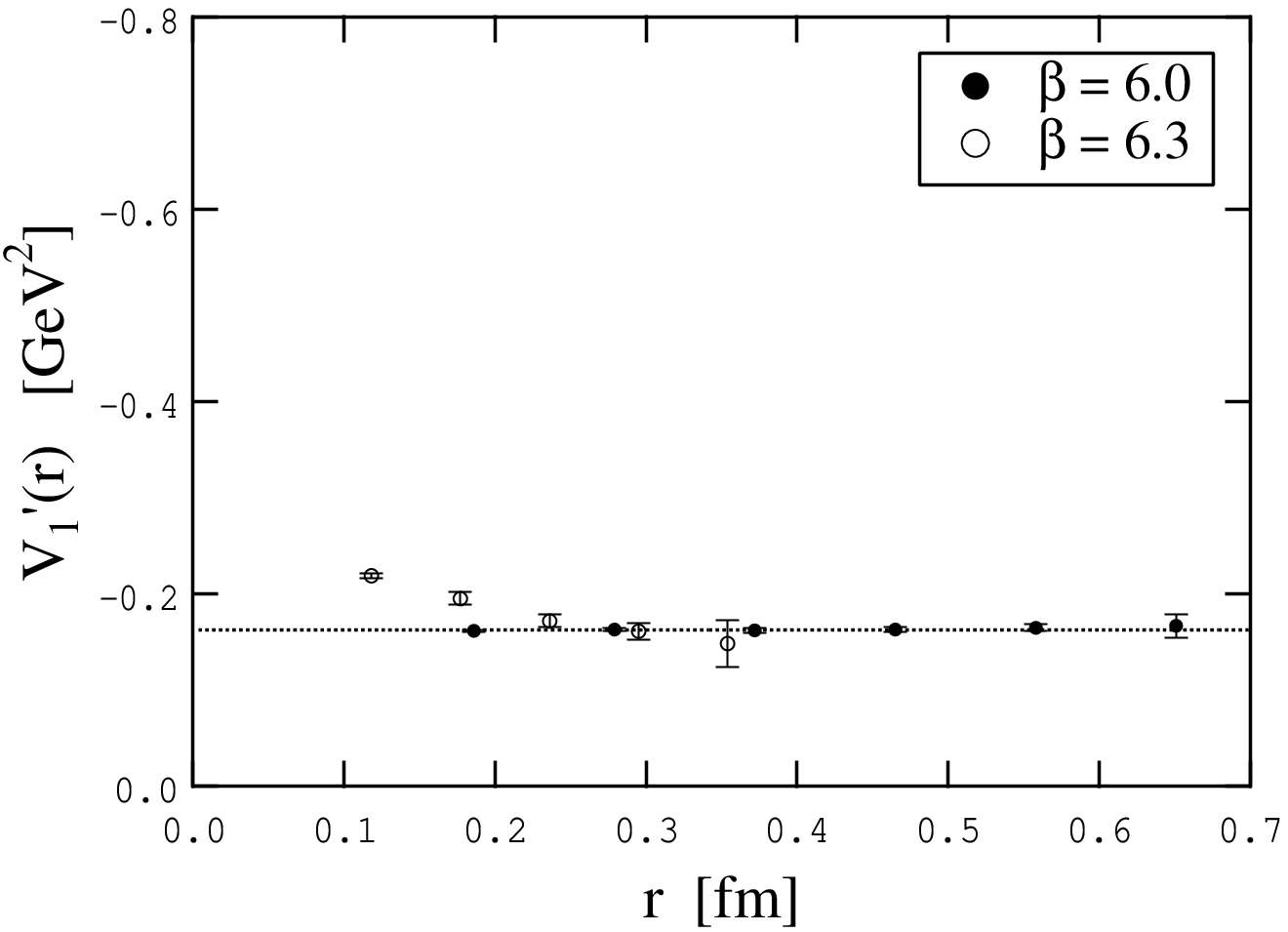}
\caption{Spin-orbit potential $V_{1}'(r)$ at $\beta=6.0$ 
and $\beta=6.3$.
The dotted line is the fit curve Eq.~\eqref{eqn:v1-fit},
applied to the data of $\beta=6.0$.}
\label{fig:v1}
\vspace*{0.5cm}
\centering
\includegraphics[width=9.5cm]{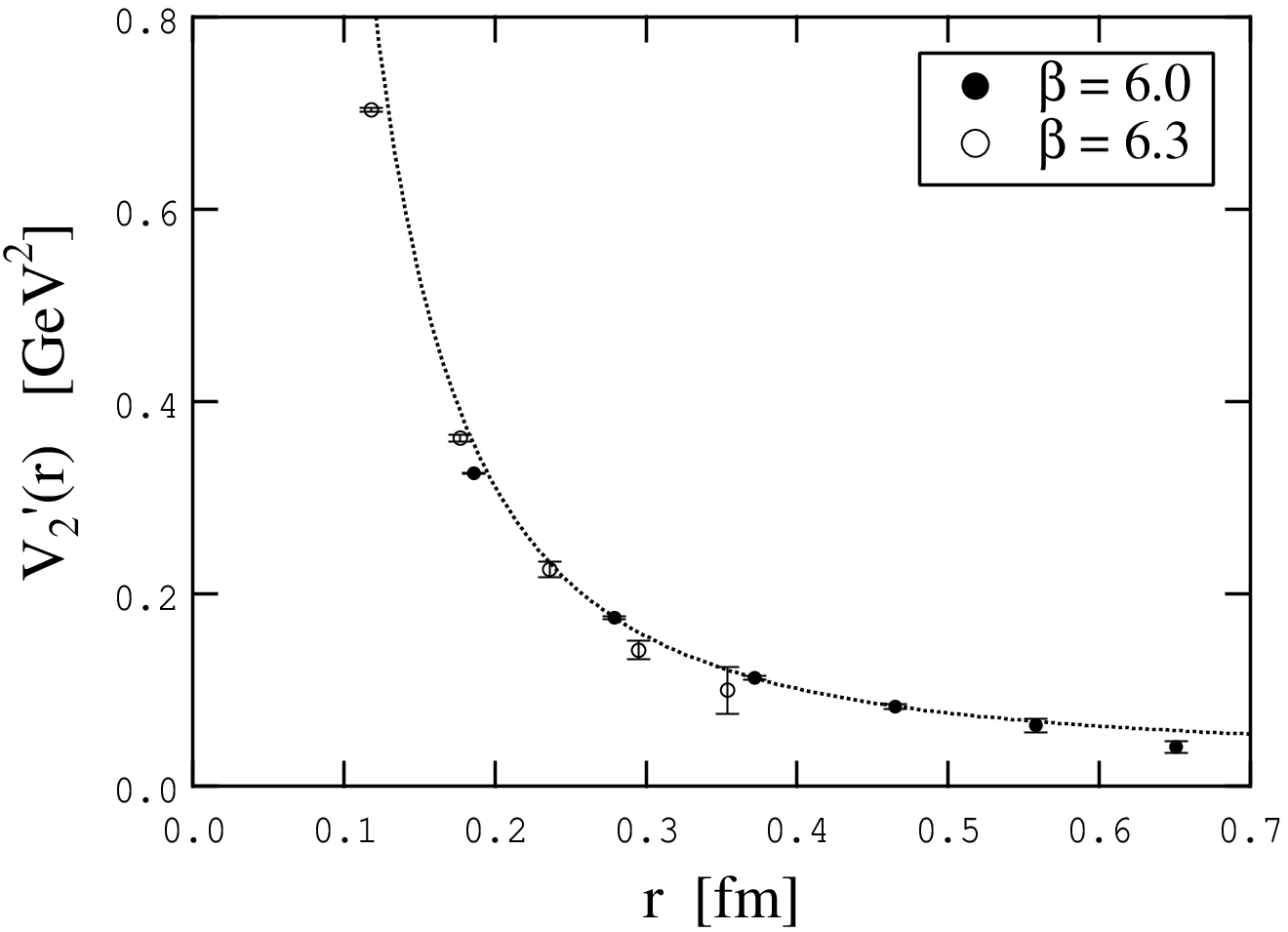}
\caption{Spin-orbit potential $V_{2}'(r)$ at $\beta=6.0$ 
and $\beta=6.3$.
The dotted line is the fit curve Eq.~\eqref{eqn:force-fit},
applied to the data of $\beta=6.0$.}
\label{fig:v2}
\end{figure}

The spin-dependent potentials,
$V_{1}'(r)$, $V_{2}'(r)$, $V_{3}(r)$ and
$V_{4}(r)$ at $\beta=6.0$ 
on the $20^{3}40$ lattice and at $\beta = 6.3$ on the
$24^4$ lattice are presented
in Figs.~\ref{fig:v1},~\ref{fig:v2},~\ref{fig:v3} 
and \ref{fig:v4}, respectively,
expressed in physical units.
These are the main results of this paper.
Though we expect a scaling behavior for $V_{\rm SD}(r)$ 
in Eq.~\eqref{eqn:potential}, 
both data at $\beta=6.0$ and $\beta=6.3$ for each potential
seem to fall into one curve,
which in turn may indicate that the matching coefficients
should depend weakly on $\beta$.
The qualitative behavior of these potentials is not
obscured by numerical errors.
However, there is still room for improvement
for the data with $r> 0.3$~fm at $\beta=6.3$.
The raw data in the lattice unit are summarized in 
Tables~\ref{tbl:spin-L20b60}-\ref{tbl:spin-L24b63}
in the Appendix.\footnote{In these data, the HM factors are already 
multiplied, but the bare lattice data can be extracted by dividing
the corresponding factors in Table~\ref{tbl:HMfactor}. 
Starting from the bare data
one can also test other renormalization procedures.}
The rest of this subsection is devoted to the interpretation of our
data. In particular, we shall discuss the functional form of the
dependence of the potentials on the distance~$r$.

\par
We start by briefly summarizing the theoretical 
expectation for the spin-dependent potentials.
As mentioned in the introduction, Gromes derived a relation
between the static potential and the spin-orbit potentials, 
$V'_{0}(r)=V'_{2}(r)-V'_{1}(r)$,
using the Lorentz (Poincar\'e) invariance of the field strength
correlators~\cite{Gromes:1984ma,Brambilla:2001xk}.
He also derived several inequalities for the spin-spin potentials
based on reflection positivity, such as
$V_{3}(r)\geq V_{4}(r)$ and 
$2V_{3}(r)+V_{4}(r)\geq 0$~\cite{Gromes:1987zx}.
These relations are nonperturbative, and 
can thus be checked directly on the lattice.\footnote{One 
may of course expect a certain deviation from this relation on the 
lattice with a finite lattice cutoff $a$,
since the strict Lorentz invariance is restored only 
in the continuum limit, $a \to 0$.}
Moreover, these relation do not depend on the matching scale.

\par
Another source of information comes from the
systematic non-relativistic reduction of the
Bethe-Salpeter (BS) equations within the instantaneous 
approximation~\cite{Lucha:1991vn}.
Starting from the interaction kernel, 
which is assumed to be a function of
the norm squared of the relative momentum between a quark
and an antiquark with various Lorentz structures,
one arrives at a Breit-Fermi type effective Hamiltonian up to 
$O(1/m^{2})$~\cite{Gromes:1976np,Gesztesy:1983bj}.
By comparing this effective Hamiltonian with
Eq.~\eqref{eqn:potential} (where $C_{F}^{(i)}=1$ is assumed), 
one obtains the relation between
the Lorentz structure of the kernel and the spin-dependent 
potentials as summarized in Table~\ref{tbl:kernel}.
This indicates that the Lorentz structure of the
confining static potential can also be inferred from 
the structure of the spin-dependent potentials.
For the special case of the one-gluon-exchange 
interaction, the kernel only consists of 
the Lorentz vector, and 
the spin-dependent potentials are explicitly given~by
\bea
V_{1}'(r)=0 \; ,\quad V'_{2}(r) = \frac{c}{ r^{2}} \; ,
\quad   V_{3}(r) =\frac{3c}{ r^{3}} \;, \quad
V_{4}(r)=8 \pi c \delta^{(3)}(r)\; ,
\label{eqn:perturbation}\quad\quad
\eea
where $c=C_{F}\alpha_{s}$.

\begin{table}[tb]
\centering
\caption{The relation between the Lorentz structure of the
effective kernel in the Bethe-Salpeter equation and the 
spin-dependent potentials~\cite{Lucha:1991vn}. 
$S(r)$, $V(r)$ and $P(r)$ 
are some scalar functions.
If the interaction kernel has several components,
the expected forms of the potentials are given by
the sum of the corresponding terms.}
\vspace{3mm}
\begin{tabular}{lccccc}
    \hline
    Kernel & $V_{0}(r)$ & $V_{1}(r)$ & $V_{2}(r)$ & $V_{3}(r)$ 
    & $V_{4}(r)$  \\
    \hline
    Scalar & $S(r)$ & $-S(r)$ & $0$ & 0 & 0  \\
    Vector & $V(r)$ & 0 & $V(r)$ & $-V''(r)+V'(r)/r$ & $2\Delta V(r)$  \\
    Pseudo-scalar & 0 & 0 & 0 &$ P''(r)-P'(r)/r$ & $\Delta P(r)$  \\
    \hline
\end{tabular}
\label{tbl:kernel}
\vspace*{0.5cm}
\end{table}

\par 
We shall now investigate the $r$-dependence of our
lattice data in more detail. 
We emphasize that, apart from the Gromes relation, no
exact predictions exist for the behavior of the
potentials beyond the short-distance regime. 
We have thus investigated the 
$r$-dependence of the potential 
by fitting the data to a particular model
function, mostly guided by the short-distance predictions of
Eq.~\eqref{eqn:perturbation}. 
In cases where the latter clearly failed
to describe the data, we have also resorted to effective
parametrizations. The quality of a particular fit ansatz 
was judged by monitoring the value of
$\chi^2_{\rm min}/N_{\rm df}$ as computed
using the full covariance matrix. 
Clearly, the underlying mechanism
responsible for the observed behavior cannot be
established rigorously in this manner.
However, the main aim of this analysis is to provide
nonperturbative input and guidance for future conceptual
studies in this area.

\par
In the following we concentrate mostly on the dataset 
at $\beta=6.0$,
since it extends to larger distances compared
with the data collected at $\beta=6.3$. 
On the other hand, at smaller distances the results
may be affected more by lattice artefacts.
Indeed, around $r\approx 0.2$\,fm we occasionally observe 
small discrepancies for some potentials.
Therefore, in the fits to the $\beta = 6.0$ dataset
described below, we have mostly omitted the data point
corresponding to the smallest separation $r/a=2$.

\par
We start our discussion with the spin-orbit potentials
$V_1'(r)$ and $V_2'(r)$.
For $V_1'(r)$, we find that the potential is negative 
and almost constant at $r \gtrsim 0.25$~fm 
(see Fig.~\ref{fig:v1}).
This behavior clearly 
contradicts Eq.~\eqref{eqn:perturbation} and
suggests the existence of the Lorentz-scalar piece in the interaction
kernel in terms of the BS equation.  
Our data at $\beta = 6.3$ suggest that one
cannot exclude a deviation from a constant at small distances, an
observation which was also made by 
Bali et al.~\cite{Bali:1996cj,Bali:1997am}.
For $V_{2}'(r)$, we see a decreasing behavior with $r$ 
(see Fig.~\ref{fig:v2}), which
is not restricted to the short range, but
rather seems to have a finite tail up to intermediate distances.

\begin{figure}[t]
\centering\includegraphics[width=9.5cm]{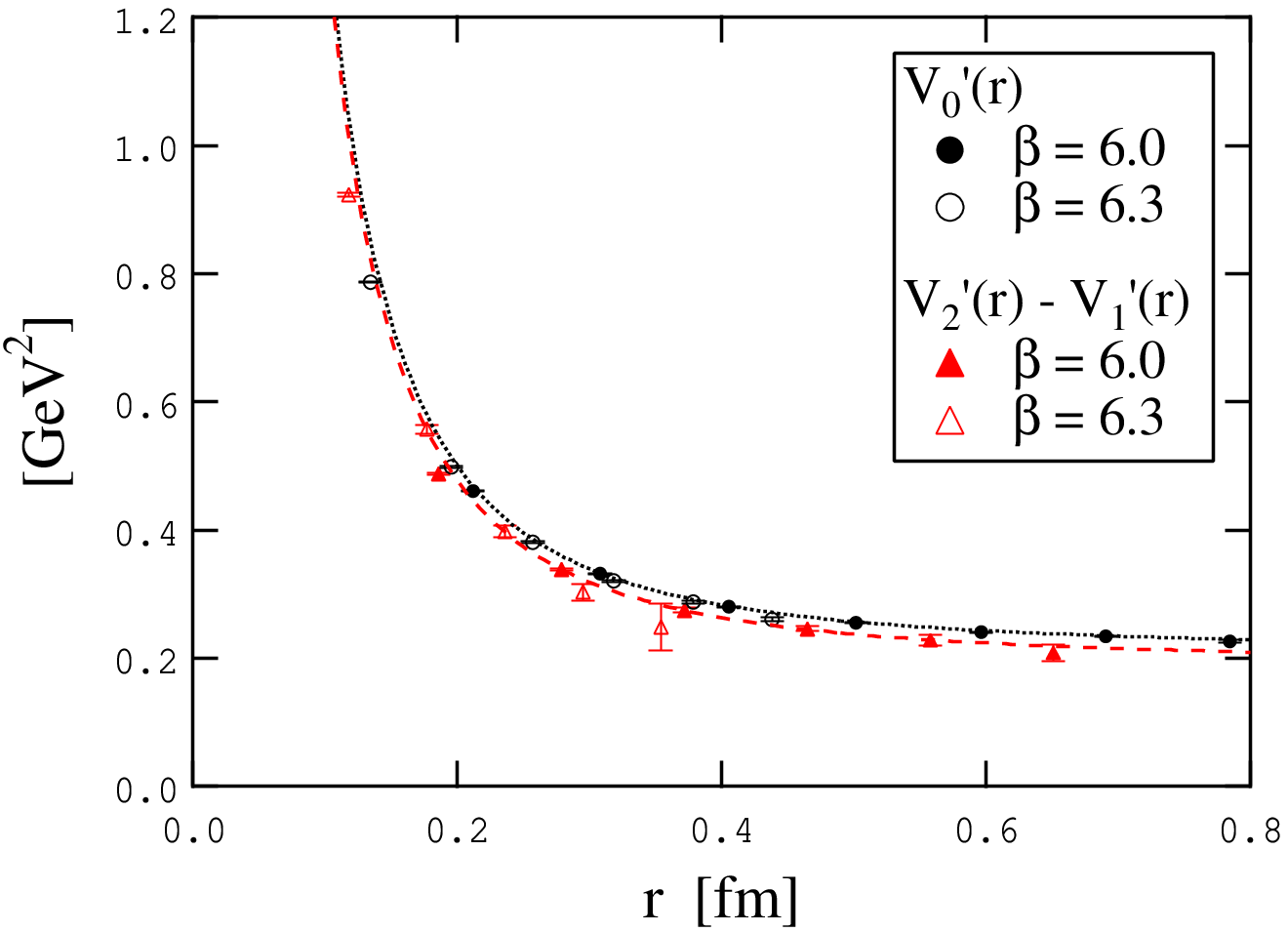}
\caption{Comparison between the force $V_{0}'(r)$ and 
the difference of the spin-orbit potentials $V_{2}'(r)-V_{1}'(r)$
at  $\beta=6.0$ on the $20^3 40$ lattice 
and $\beta=6.3$ on the $24^4$ lattice
with the physical scale.
The Gromes relation is $V_{0}'(r)=V_{2}'(r)-V_{1}'(r)$.
The dotted line is the fit curve for the force $V_{0}'$, while 
the dashed line is for $V_{2}'-V_{1}'$,
where Eq.~\eqref{eqn:force-fit} is used in both cases.}
\label{fig:grms_sc}
\vspace*{0.4cm}
\centering\includegraphics[width=9.5cm]{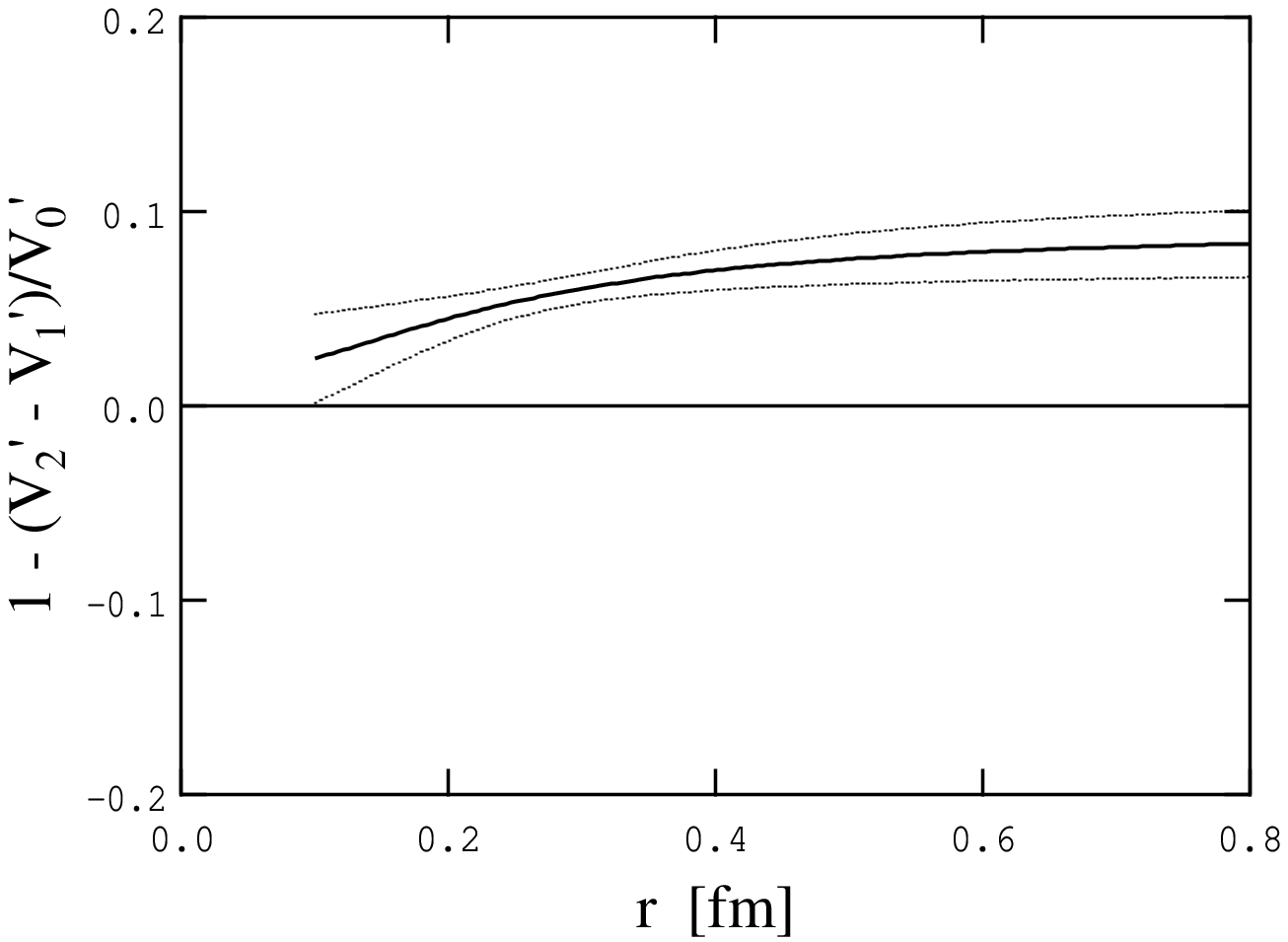}
\caption{The relative deviation from the Gromes relation, $1-
  (V_{2}'-V_{1}')_{\rm fit}/ V_{\rm 0;fit}'$, at
  $\beta=6.0$. The fit functions are
  the same as in~Fig.~\ref{fig:grms_sc} and the
dotted lines denote the $1\sigma$ error band.}
\label{fig:force_grms_sc}
\vspace*{0.4cm}
\end{figure}

\par
Before discussing the functional form of $V_{1}'(r)$ and 
$V_{2}'(r)$ quantitatively,
we may ask if these spin-orbit potentials
satisfy the Gromes relation, since otherwise
it is unclear whether any quantitative 
arguments make sense. 
In~Fig.~\ref{fig:grms_sc}, we compare the static
force, $V_{0}'(r)$,
with the difference of the spin-orbit potentials,
$V_{2}'(r)-V_{1}'(r)$.
We find quite a good agreement, indicating that
the Gromes relation is apparently satisfied.
We can examine this relation in more detail by
fitting the difference $V_{2}'(r)-V_{1}'(r)$ at $\beta = 6.0$ 
to the same functional form as the force
in Eq.~\eqref{eqn:force-fit}.
Thereby we obtain $\sigma_{\rm v21} a^2 = 0.0426(9)$
and $c=0.293(9)$ with $\chi^2_{\rm min}/N_{\rm df}=0.26$.~\footnote{Here
and in the following, we attach a subscript to $\sigma$ so as to distinguish
the target function in the fit.}
The string tension extracted in this way is about
8~\% smaller than that in $V_{0}'(r)$,
while the Coulombic coefficients are in agreement within errors.
We also plot the quantity $1-(V_{2}' -V_{1}')_{\rm fit}/V_{\rm 0;fit}'$ in 
Fig.~\ref{fig:force_grms_sc}. From 
this we conclude that the Gromes relation is satisfied
within $(8 \pm 1)$~\% accuracy at $r \approx 0.5$~fm.
Note that without the renormalization factor for the field strength operator,
one would observe a strong deviation 
from the Gromes relation by a factor $\approx 2.7$ at $\beta=6.0$.
In this sense, the renormalization of the operator  is crucial 
for satisfying the Gromes relation within a few percent level,
especially when the lattice spacing is finite.
It is certainly interesting to investigate 
if the Gromes relation is exactly satisfied
in the continuum limit.
Although we have investigated the gauge coupling at 
$\beta=6.3$, we need further accuracy of the data at 
intermediate distances to achieve this.

\par
For $V_{1}'(r)$, 
if we only take into account the data for $r\gtrsim 0.25$~fm
at $\beta=6.0$ and, assuming that they are constant, 
we can fit them to a function 
\bea
V'_{\rm 1;fit} = -\sigma_{\rm v1} \; .
\label{eqn:v1-fit}
\eea
Due to the Gromes relation, we may identify this constant
as a part of the string tension in $V_{0}'(r)$.
We then find $\sigma_{\rm v1}a^2 = 0.0362(4)$
with $\chi^2_{\rm min}/N_{\rm df}=0.13$,
which is  $(77 \pm 1)$~\% of the string tension in $V_{0}'(r)$.
The corresponding fit curve is plotted in Fig.~\ref{fig:v1}.

\par
While the Gromes relation is approximately satisfied, 
we find that the string tension $\sigma_{\rm v1}$ is not yet 
sufficient to reproduce the string tension $\sigma_{\rm v21}$.
In other words there is still a missing amount of the string tension.
We then notice that this must be supplied by $V_{2}'(r)$.
A fit of $V_{2}'(r)$ to Eq.~\eqref{eqn:force-fit} indeed 
leads to $\sigma_{\rm v2}a^2 = 0.0070(7)$ and $c=0.288(7)$
with $\chi^2_{\rm min}/N_{\rm df}=0.22$.
The fit curve is plotted in Fig.~\ref{fig:v2}.
Now the sum of $\sigma_{\rm v1}$ and $\sigma_{\rm v2}$
reproduces $\sigma_{\rm v21}$.
These findings suggest the existence of a long-ranged 
contribution in $V_{2}'(r)$ whose
magnitude is about one-fifth of $\sigma_{\rm v1}$,
which is  $(15 \pm 2)$~\% of the string tension in $V_{0}'(r)$.
We also attempted a fit with the expectation from
perturbation theory, by simply fixing the string tension
to be zero in the above fit.
In this case the fit clearly fails, since
$\chi^2_{\rm min}/N_{\rm df}=44$. From 
the phenomenological point of view,
one might prefer a simple parametrization like 
$\sigma = \sigma_{\rm v1}$ and 
$\sigma_{\rm v2}=0$~\cite{Buchmuller:1981fr}, but
the results obtained here slightly differ from this expectation.
We wish to point out, though, that $V'_2(r)$ 
should further be investigated at
distances larger than 0.7\,fm, in order to 
corroborate a non-vanishing
value of $V'_2(r)$ for $r \to \infty$.
We may note that Eq.~\eqref{eqn:force-fit} is not the only
functional form for $V_{2}'(r)$.
For instance, the function $V_{\rm 2; fit}'(r)= c'/r^p$
with $p=1.51(4)$ and $c' a^{p-2}=0.205(9)$ also
reproduces the data quite well,
with $\chi^2_{\rm min}/N_{\rm df}=0.34$.

\begin{figure}[t]
\centering
\includegraphics[width=9.5cm]{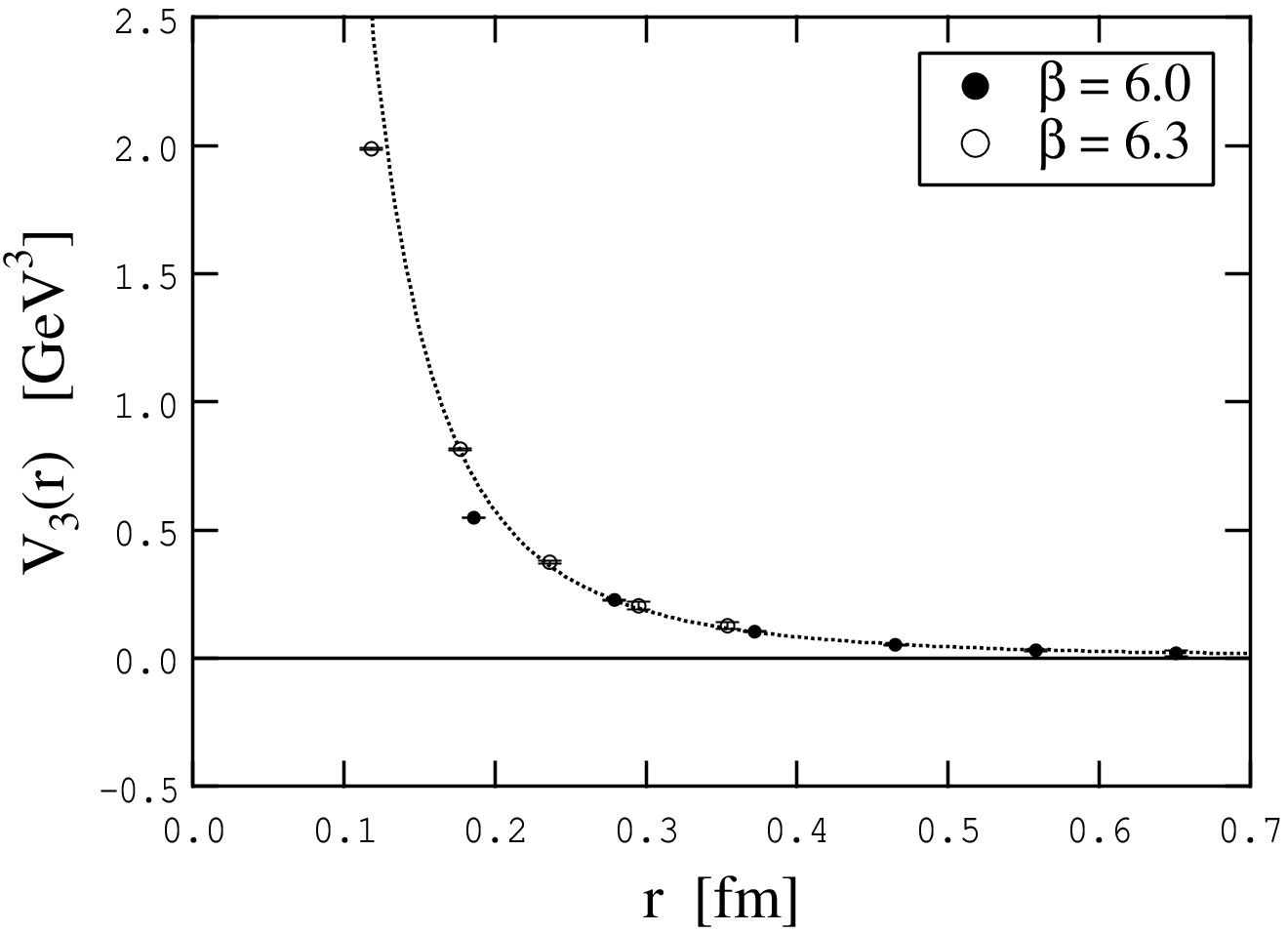}%
\caption{Spin-spin (tensor) potential $V_{3}(r)$ at $\beta=6.0$ 
and $\beta=6.3$.
The dotted line is the fit curve Eq.~\eqref{eqn:v3-fit},
applied to the data of $\beta=6.0$.}
\label{fig:v3}
\vspace*{0.4cm}
\centering
\includegraphics[width=9.5cm]{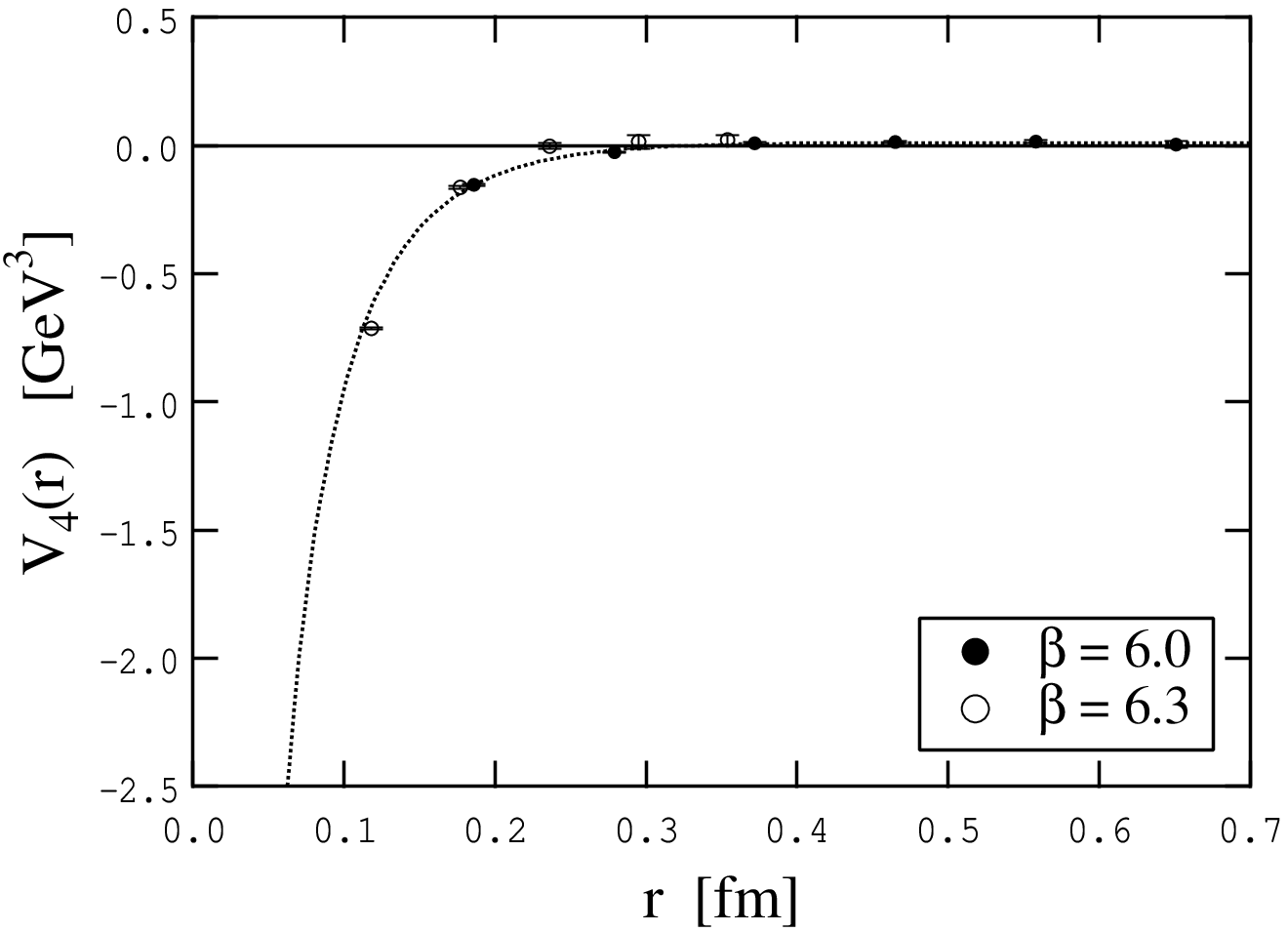}
\caption{Spin-spin potential $V_{4}(r)$ at $\beta=6.0$ 
and $\beta=6.3$.
The dotted line is the fit curve Eq.~\eqref{eqn:v4-fit},
applied to the data of $\beta=6.0$.}
\label{fig:v4}
\vspace*{0.4cm}
\end{figure}

\par
Next, we discuss the spin-spin potentials $V_{3}(r)$ and $V_{4}(r)$.
We first examine $V_{3}(r)$ (see Fig.~\ref{fig:v3}) 
if the ansatz motivated by one-gluon-exchange
in Eq.~\eqref{eqn:perturbation} is appropriate.
The fit to this function yields the coefficient $c = 0.214(2)$ 
with $\chi_{\rm min}^{2}/N_{\mathrm{df}}=3.7$.
This value of $\chi_{\rm min}^2/N_{\mathrm{df}}$ is relatively
large and the result for 
$c$ is 28 \% smaller than the Coulombic 
coefficient in $V_{0}'(r)$.
A better fit can be achieved using an ansatz in which the 
power of $1/r$ is left as a free parameter, i.e.
\bea
V_{\rm 3;fit}(r)=\frac{3c'}{r^p} \;.
\label{eqn:v3-fit}
\eea
This yields
$c' a^{p-3} = 0.171(10)$ and $p=2.80(6)$, where
$\chi_{\rm min}^{2}/N_{\mathrm{df}}=0.79$.
The corresponding fit curve is plotted 
in Fig~\ref{fig:v3}.
The value of $p$ is smaller than $3$ within 3 standard deviations. If
one takes this result as face value, it indicates a
deviation from the one-gluon-exchange potential.
A deviation might actually be expected from the 
existence of the long-ranged contribution in $V_{2}'(r)$
and the relations in Table~\ref{tbl:kernel};
if we insert a function 
$V(r)=-c/r + \sigma_{\rm v2}r$ into $-V''(r)+V'(r)/r$,
we obtain $3 c /r^{3} + \sigma_{\rm v2}/r$ at $r \ne 0$.
We have then examined if this function describes
the data for given $c$ and $\sigma_{\rm v2}$, 
which are supplied from the $V_{2}'(r)$ fit.
However, we have found that the resulting curve is  
not appropriate to describe the behavior of the data at all, 
since it lies above the data points at small distances.
This tendency is practically due to the term $1/r^{3}$,
but this additional term $1/r$ also helps to lift the curve.
It suggests that we need to add a negative contribution
to such an ansatz.

\par
A possible candidate would then be a 
pseudo-scalar contribution, which is also closely related
to the behavior of 
$V_{4}(r)$ (see Fig.~\ref{fig:v4}).
In fact, if only the one-gluon-exchange interaction is
considered in the vector kernel, 
$2 \Delta V (r)=2(V''(r)+2V'(r)/r)$ 
leads to a $\delta$-function as in Eq.~\eqref{eqn:perturbation},
while if we insert the empirical behavior of $V_{2}'(r)$,
an additional term of $4 \sigma_{\rm v2}/r$ is generated for $V_4(r)$.
Thus we expect a positive behavior at non-zero distances.
By contrast, the data is negative 
at small distances and almost zero for $r>0.2$~fm.
Let us now assume the presence of a pseudo-scalar
interaction, $P(r) = - g' e^{-m_{g}r}/r$, 
where $m_{g}$ is the mass of the 
lightest pseudo-scalar particle, and $g'$ is the
corresponding effective coupling to quarks.
This certainly generates a negative contribution,
$\Delta P(r)  =  - g' m_{g}^2 e^{-m_{g}r}/r$, to $V_4(r)$.
Note that the pseudo-scalar interaction $P(r)$ is often used
in the one-boson-exchange model
for describing the nucleon-nucleon system,
where pions play a relevant role~\cite{Gross:1991pm}.
In our simulation, however, since the effects of
dynamical fermions are
neglected due to our use of the quenched
approximation, the lowest mass in the pseudo-scalar channel
cannot be identified with the pion mass
but rather with the lightest glueball mass.

\par
We have then performed a fit to 
\bea
V_{\rm 4; fit}(r)=- g' m_{g}^2 \frac{e^{-m_{g}r}}{r} 
+4 \frac{\sigma_{\rm v4}}{r}  \; ,
\label{eqn:v4-fit}
\eea
where we have assumed $m_{g}= 2.47$~GeV, which is taken from
the recent lattice studies 
of the glueball masses~\cite{Chen:2005mg}, and
treated $g'$ and $\sigma_{\rm v4}$ as free parameters.
The result was $g'=0.292(12)$ 
and $\sigma_{\rm v4} a^2 =0.0015(3)$
with $\chi_{\rm min}^{2}/N_{\mathrm{df}}=5.1$,
and the corresponding curve is put in Fig.~\ref{fig:v4}
(if $m_{g}$ is relaxed to be a free parameter,
$\chi_{\rm min}^2/N_{\rm df}$ is significantly reduced).
We find that $\sigma_{\rm v4}$ (notice that this value is not zero)
is not exactly $\sigma_{\rm v2}$, but as the relation among interaction 
kernels is not  exact but derived within the
instantaneous approximation, such a deviation may occur,
especially for the nonperturbative pieces.
On the other hand, the value of $g'$ is very close to
$c =0.297(1)$ determined from the 
force (see Table~\ref{tbl:potfit}).
If we impose $\sigma_{\rm v4}=\sigma_{\rm v2}$,
we obtain here
$\chi_{\rm min}^{2}/N_{\mathrm{df}}\approx 100$.
However the apparent shape of the curve
is not affected so much as is mostly governed 
by the glueball mass we set.

\begin{figure}[t]
\centering\includegraphics[width=9.5cm]{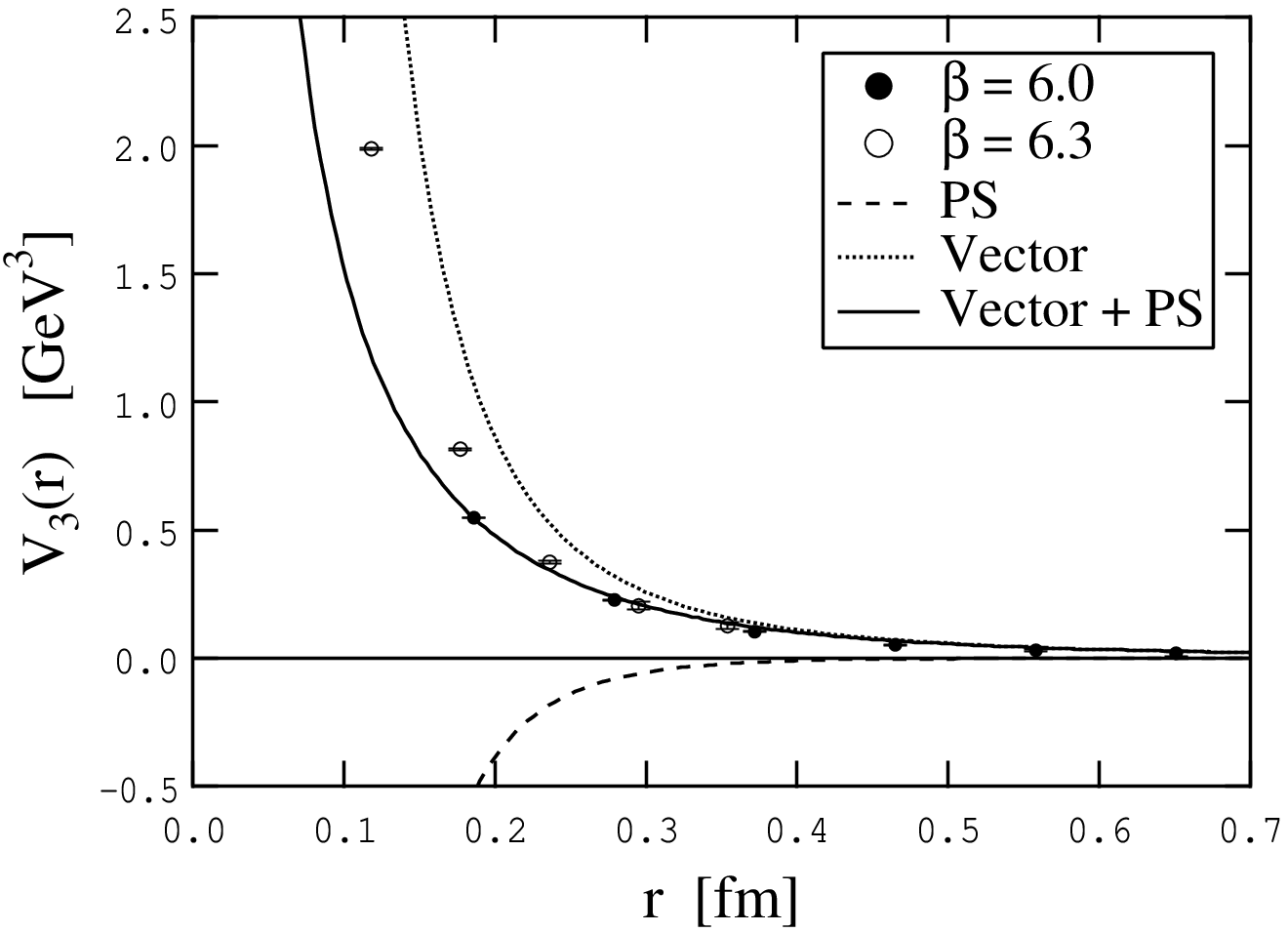}
\caption{A possible structure of the spin-spin potential $V_{3}(r)$.
The lattice data are the same as in Fig.~\ref{fig:v3}.
The dotted (dashed) line corresponds to the vector 
(pseudo-scalar) contribution, 
Eq.~\eqref{eqn:v3-vector} (Eq.~\eqref{eqn:v3-pscalar})
and the sum of these two
contributions is indicated by the solid line.}
\label{fig:v3_ps}
\vspace*{0.5cm}
\centering\includegraphics[width=9.5cm]{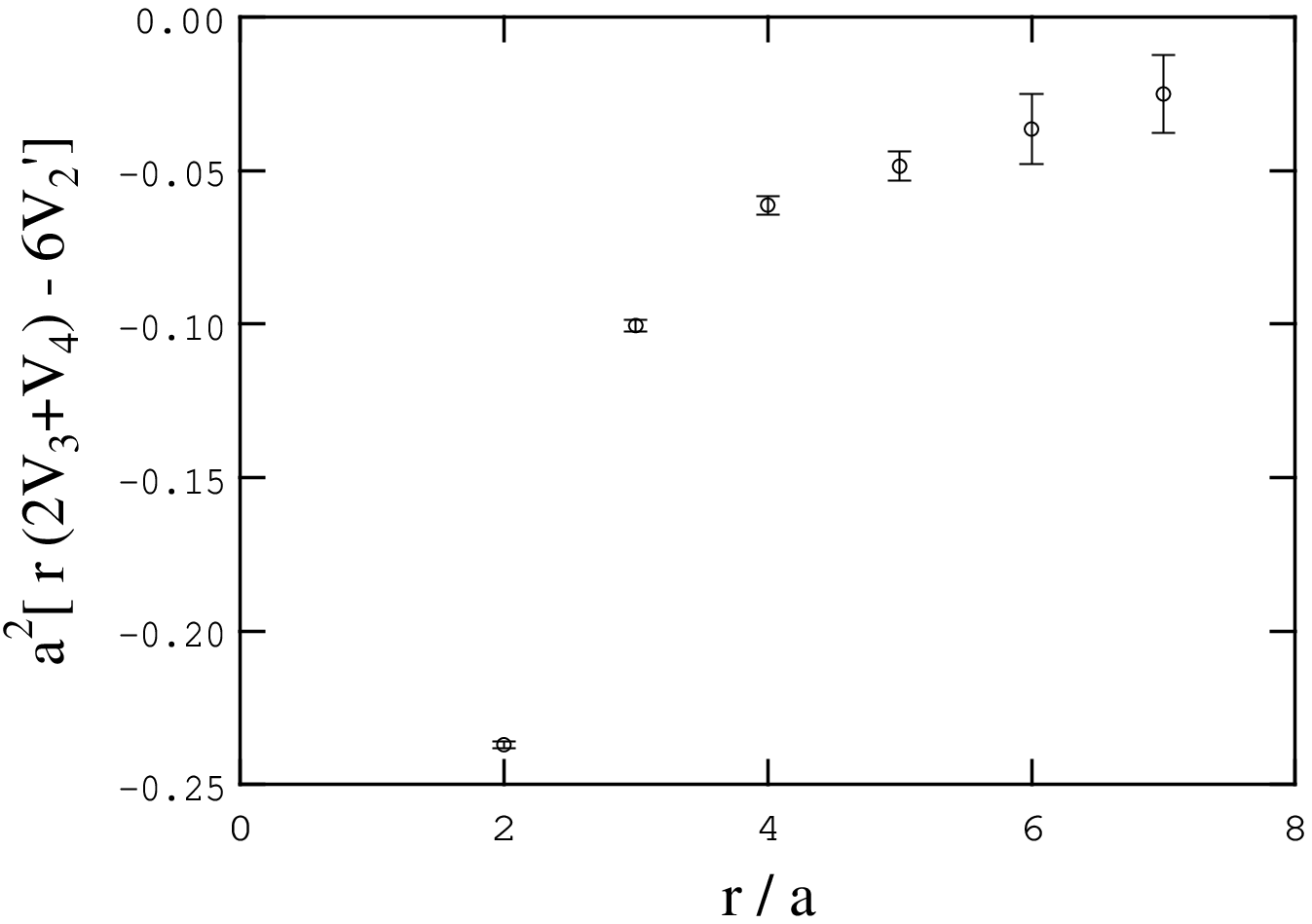}
\caption{$r (2V_{3}+V_{4}) -6V_{2}'$ at $\beta=6.0$.}
\label{fig:r2v3v4-6v2}
\vspace*{0.5cm}
\end{figure}

\par
Now we come back to $V_{3}(r)$ and
examine if the sum of the (positive) vector and the (negative) 
pseudo-scalar contributions, with parameters estimated 
by the $V_{4}(r)$ fit, can describe the behavior of $V_{3}(r)$.
Here, we consider the sum of two functions,
\bea
&&V_{3}^{\rm (V)}(r) = \frac{3 c }{r^3}
+ \frac{\sigma_{\rm v4}}{r} \; ,
\label{eqn:v3-vector}\\
&&V_{3}^{\rm (PS)}(r) =  P''(r)- \frac{P'(r)}{r} 
= - g'(\frac{3}{r^{3}} +\frac{3m_{g}}{r^{2}}
+\frac{m_{g}^2}{r}) e^{-m_{g} r} \; ,
\label{eqn:v3-pscalar}
\eea
where $c$ is taken from $V_{0}'(r)$.
It is interesting to see Fig.~\ref{fig:v3_ps}
that the curve $V_{3}^{\rm (V)}(r)+V_{3}^{\rm (PS)}(r)$,
which is plotted with the solid line,
can go through the data at $\beta=6.0$.

\begin{table}[t]
\centering
\caption{Fit results of the spin-dependent potentials
at $\beta=6.0$ on the $20^3 40$ lattice.
(*) if we relax $m_{g}$ to be a free parameter,
$\chi_{\rm min}^{2}/N_{\mathrm{df}}$ is 
significantly reduced.}
\vspace*{0.3cm}
\begin{tabular}{lclc}
\hline
Potential & Fit range ($r/a$) & Fit function and parameters & 
$\chi_{\rm min}^{2}/N_{\mathrm{df}}$\\
\hline
$V_{1}'(r)$  
&  & $V_{\mathrm{fit}}' =   -\sigma$ &\\
& $3 - 7$ & $\sigma a^2 = 0.0362(4)$  & 0.13 \\
\hline
$V_{2}'(r)$  
&  &$V_{\mathrm{fit}}'(r) =   \sigma +c/r^2$ &\\
& $3 - 6$ & $\sigma a^2 = 0.0070(7)$, $c=0.288(7)$  & 0.22 \\
\cline{2-4}
&  & $V_{\mathrm{fit}}'(r) =  c'/ r^{p}$ & \\
 & $3 - 6$ & $c' a^{p-2} = 0.205(9)$, $p=1.51(4)$  & 0.34 \\
\hline
$V_{2}'(r)-V_{1}'(r)$  
&  &$ V_{\mathrm{fit}}'(r) =   \sigma +c/r^2$ &\\
& $3 - 7$ & $\sigma a^2 = 0.0426(9)$, $c=0.293(9)$  & 0.26 \\
 \hline
$V_{3}(r)$ 
&  &$ V_{\mathrm{fit}}(r) =  3c/ r^{3}$ & \\
 & $3 - 7$ & $c=0.214(2)$  & 3.7{\F}\\
 \cline{2-4}
&  &$ V_{\mathrm{fit}}(r) =  3c'/ r^{p}$ & \\
 & $3 - 7$ & $c' a^{p-3}=0.171(10)$, $p=2.80(6)$  & 0.79\\
\hline
$V_{4}(r)$ 
&  &$ V_{\mathrm{fit}}(r) =  
- g' m_{g}^2 e^{-m_{g}r}/r +4 \sigma/r $ & \\
& $2 - 7$ & $g'=0.292(12)$, $\sigma a^2= 0.0015(3)$, &   \\
 &  & $m_{g}a=1.16$ (fixed)   & 5.1$^{*}$\\
 \hline
\end{tabular}
\label{tbl:spinpot-fit}
\vspace*{0.5cm}
\end{table}

\par
In most of previous works, it was concluded
that $V_{4}(r)$ is consistent with a
$\delta$-function,
which may only be true at very small distances.
However, as we demonstrated,  the behavior of $V_{3}(r)$
and $V_{4}(r)$ at distances $r\gtrsim 0.2$~fm can be
consistently explained by assuming the existence of
the pseudo-scalar contribution as well as the vector contribution.
Note furthermore that the combination of the
potentials $r(2V_{3}+V_{4})-6V_{2}'$
should be zero at non-zero distances,
if the interaction kernel
contains only the pure vector component 
without a linear term and  no pseudo-scalar contribution 
(see Table~\ref{tbl:kernel})~\cite{Huntley:1986de}.
However, as shown in Fig.~\ref{fig:r2v3v4-6v2},
we find that this combination is non-vanishing within our 
accuracy, so that some of these assumptions are probably
not applicable.
Of course, our discussion on the pseudo-scalar contribution
is as yet speculation, 
which needs to be checked in future works.
A possible way of doing this is
to investigate $V_{4}(r)$ in the presence of 
dynamical quarks (pions) in full QCD simulations,
and to examine whether one can indeed observe a behavior like 
$\propto - e^{-m_{\pi}r}/r$ for sufficiently small quark masses.
Some of previous works in Refs.~\cite{Koike:1989jf,Born:1993cp}
have been carried out in full QCD, but 
the data quality is not sufficient to draw any conclusion.
In any case, we expect to raise further discussions on 
the structure of the spin-spin potentials.

\par
To close the discussion on the functional form,
we note that the Gromes inequalities
$V_{3}(r)\geq V_{4}(r)$ and $2V_{3}(r)+V_{4}(r) \geq 0$
are certainly satisfied.
For instance, the latter inequality is immediately checked
through $2V_{3}(r)+V_{4}(r) =
6 \int_{0}^{\infty} dt \[ g^2 B_{x}(\vec{0},0)B_{x}(\vec{r},t)\]$,
which is positive at all available $r$ as can be seen from 
Tables~\ref{tbl:corfit-L20b60}-\ref{tbl:corfit-L24b63}
in the Appendix.
We summarize all fit results of the functional form
in Table~\ref{tbl:spinpot-fit}.

\section{Summary}
\label{sec:summary}

\par
We have investigated the spin-dependent corrections to the 
static potential at $O(1/m^2)$ in SU(3) lattice gauge theory.
These corrections,  usually called the spin-dependent potentials,
are represented as the integral of the 
field strength correlators on the quark-antiquark source
with respect to the relative temporal distance between two field
strength operators.
We have used the Polyakov loop correlation function
as the  quark-antiquark source, and
by employing the multi-level algorithm,
we have obtained remarkably clean data
for the expectation values of the field strength correlators
and, in turn, for the spin-dependent potentials
up to intermediate distances of around $r \simeq 0.6$~fm.
The spectral  representation of the field strength correlator 
in a finite periodic volume has been exploited 
in order to extract the potential with less systematic error.

\par
The observation we have made for the spin-dependent 
potentials in Eq.~\eqref{eqn:potential} is as follows.
The spin-orbit potential $V_{1}'(r)$ is clearly 
long-ranged, is negative  at all 
distances and constant at $r \gtrsim 0.25$~fm.
The other spin-orbit potential $V_{2}'(r)$ is positive at all 
distances and shows a behavior decreasing with~$r$.
However, it has a finite tail up to intermediate distances,
which cannot be explained at least by the one-gluon-exchange
interaction.
The Gromes relation $V_{0}'(r) =V_{2}'(r)-V_{1}'(r)$ is
satisfied within ($8 \pm 1$)~\% accuracy at intermediate 
distances in the present simulation.
Within this relation, the constant value in $V_{1}'(r)$
reproduces  $(77 \pm 1)$~\% 
of the string tension in $V_{0}'(r)$
and $(15 \pm 2)$~\% of the string tension are
found to be supplied by $V_{2}'(r)$.
The spin-spin (tensor) potential $V_{3}(r)$ is positive  at all 
distances and is decreasing as a function of~$r$.
The behavior is slightly more moderate than the expectation 
of the one-gluon-exchange picture~$\propto 1/r^{3}$.
The other spin-spin potential $V_{4}(r)$ exhibits
a negative short-ranged behavior.
This short-ranged behavior, as well as the behavior of $V_{3}(r)$,
could be explained if  the exchange of the pseudo-scalar 
glueball is assumed in addition to the one-gluon-exchange 
type interaction.

\par
In this paper we have not carried out a detailed comparison
of the lattice result of the spin-dependent potentials 
with perturbation theory, e.g. along the lines of
Necco and Sommer for the static 
potential~\cite{Necco:2001xg,Necco:2001gh}.
In this sense, although we have observed a certain 
deviation from the expectation of 
leading order perturbation theory at intermediate distances,
it is not yet clear that from which distance 
a perturbative description becomes inadequate.
Clearly, it requires further systematic studies,
where the renormalization of the field strength operator
and also the matching coefficients are worth to be reconsidered.
However, we expect that the numerical procedures
we have demonstrated in this paper is
quite useful for such a work.
Then, it is interesting to use the result
as inputs of phenomenological 
models~\cite{Ebert:1997nk,Ebert:1999xv,Ebert:2002pp}
or to compare with the various QCD vacuum
models~\cite{Simonov:1988mj,Brambilla:1996aq,Jugeau:2003df}.
It may be interesting to note that the existence of a
long-ranged contribution in $V_{2}'(r)$ is suggested 
in Ref.~\cite{Jugeau:2003df}, independently of 
the present work.

\par
Finally we note that our numerical procedures are also applicable
to the evaluation of other relativistic corrections 
like the velocity-dependent 
potentials~\cite{Barchielli:1986zs,Barchielli:1988zp,Pineda:2000sz}
and the potential at 
$O(1/m)$~\cite{Brambilla:1999xf,Brambilla:2000gk,Pineda:2000sz,%
Melnikov:1998pr,Hoang:1998uv,Brambilla:1999xj,Kniehl:2001ju,%
Kniehl:2002br},
which are also represented as the field strength correlators 
on the quark-antiquark source with different combination 
of the field strength operators from the spin-dependent  potentials.
The first lattice result on the potential at 
$O(1/m)$ was published in Ref.~\cite{Koma:2006si}.

\section*{Acknowledgments}

\par
We are indebted to H.~Wittig for many critical discussions and 
comments on the manuscript.
We wish to thank R.~Sommer, Ph.~de~Forcrand, 
N.~Brambilla, A.~Vairo, A.~Pineda, S.~Sint, 
D.~Ebert, V.O.~Galkin, R.N.~Faustov, P.~Weisz 
for useful discussions,
and W.~Buchm\"uller for introducing 
to us Ref.~\cite{Buchmuller:1992zf}.
We also thank G.~Bali for his correspondence and fruitful discussions.
The main calculation has been performed on the NEC SX5 
at Research Center for Nuclear Physics (RCNP), 
Osaka University, Japan.
We thank H.~Togawa and A.~Hosaka for technical supports.

\clearpage
\appendix
\section{Collection of numerical values}

\subsection{Tree-level improvement of the quark-antiquark
distances}

\begin{table}[hbt]
\centering
\caption{The quark-antiquark distances 
for the static potential $V_{0}(r_{I})$, the force $V_{0}'(\bar{r})$ 
(or $V_{0}'(\bar{r}_{c})$)
and the second derivative $V''_{0}(\tilde{r})$ with the tree-level 
improvement~\cite{Sommer:1993ce,Necco:2001xg,Luscher:2002qv}.}
\vspace{3mm}
\begin{tabular}{cccccc}
\hline
$r/a$ &  $r_{I}/a$  & $\bar{r}/a$ & $\bar{r}_{c}/a$ & $\tilde{r}/a$ \\
\hline
1 & 0.925  &           &   &  \\
2 & 1.855  &   1.358        & 1.649 & 1.788 \\
3 &2.889  &   2.277         & 2.654 & 2.700  \\
4 & 3.922  &  3.312   & 3.729 & 3.729 \\
5  &4.942 &  4.359   &  4.794 & 4.786 \\
6  & 5.954 &5.393     &  5.837 & 5.833  \\
7 & 6.962 &6.414      & 6.865 & 6.864  \\
8 &7.967 & 7.428     & 7.885 & 7.886  \\
9  &8.971 &8.438      & 8.899  & 8.901  \\
\hline
\end{tabular}
\label{tbl:distance}
\end{table}

\subsection{HM factors}

\begin{table}[hbt]
\centering
\caption{The HM renormalization factors at $\beta=6.0$ 
on the $20^4$ lattice (upper) and $\beta=6.3$ 
on the $24^4$ lattice (lower), where the quark-antiquark
system is set along the $x$ axis.
Thus, one should observe 
$Z_{E_{y}}=Z_{E_{z}}$, $Z_{B_{y}}=Z_{B_{z}}$.}
\vspace{3mm}
\begin{tabular}{cllllll}
\hline
$r/a$  &   $Z_{E_{x}}$&   $Z_{E_{y}}$&   $Z_{E_{z}}$
&   $Z_{B_{x}}$&   $Z_{B_{y}}$ &   $Z_{B_{z}}$ \\
\hline
   2  & 1.59446(4)     &  1.63031(6)    &  1.63038(5) &
    1.67833(16) &  1.67614(12)  & 1.67600(15)\\
   3  & 1.61170(4)   &  1.62498(6)   &  1.62503(5)  &
   1.67764(16)&   1.67661(12) &1.67651(15)  \\
   4  & 1.61620(4)   &  1.62338(6)   &  1.62339(6) &
   1.67735(16)&   1.67676 (12)&   1.67669(14)  \\
   5  & 1.61777(6)    &  1.62282(6)   &  1.62282(7) &
    1.67726(16)&  1.67687(13)  & 1.67678(15)  \\
   6  & 1.61846(6) &  1.62250(7)    &  1.62262(6) &
    1.67721(16)&   1.67695(13) &  1.67683(16)\\
   7  & 1.61877(10)   &  1.62246(8)   &  1.62233(8)& 
    1.67726(16)&  1.67684(13)  &1.67680(15)  \\
   8  & 1.61879(15)   &  1.62225(17)   &  1.62232(16)& 
   1.67708(18)&   1.67682(19) &  1.67674(19)  \\
\hline
   2  & 1.54232(3)  &1.56529(3)   &  1.56526(3) &
   1.60307(7)  & 1.60179(9) &  1.60185(7)\\
   3  & 1.55417(5) &  1.56151(4) &1.56154(4)  &
   1.60271(7)  &  1.60217(10) & 1.60226(7)\\
   4  & 1.55717(6) &1.56048(4)   &   1.56049(6)   &
   1.60266(7)    &   1.60225(10) &   1.60235(7)\\
   5  & 1.55810(9)   & 1.56013(6)   &  1.56004(7)   &
    1.60260(9)  &  1.60231(11) &  1.60231(10)  \\
   6  & 1.55853(10) &   1.55974(11)  &   1.55986(13)  &
   1.60247(11)  &   1.60229(12)   &  1.60253(11)  \\
   7  & 1.55829(18) &   1.55921(23)  &   1.55980(14)   &
    1.60238(18)  &   1.60216(17)   &   1.60223(16)   \\
   8  & 1.55855(40)  &   1.55974(37)&   1.55975(32) &
   1.60260(29)  &   1.60249(28)  &   1.60217(23) \\
   9  & 1.55961(43) &   1.55954(64)  &  1.56021(48)   &
   1.60324(36)  &  1.60324(39)  &   1.60331(44)  \\
\hline
\end{tabular}
\label{tbl:HMfactor}
\end{table}

\clearpage
\subsection{Fit results of the field strength correlators}

\begin{table}[hbt]
\caption{Fit results of the field strength correlators
with Eqs.~\eqref{eqn:spectral-rep1}-\eqref{eqn:spectral-rep3}
at $\beta=6.0$ on the $20^4$ lattice.
$m_{\rm max}$ is the maximum truncation level of
the spectral representation.}
\vspace{3mm}
\centering
\begin{tabular}{cccclc}
\hline
$C(t)$ & $r/a$ & Fit range $(t/a)$ & $m_{\rm max}$ & 
$a^2\int_{0}^{\infty} dt \; t C(t)$ & $\chi_{\rm min}^2/N_{\rm df}$ \\
\hline
$\[g^2 B_{y}(0,0) E_{z}(0,t) \]$
&  2  & $2-10$   &2      &         $-$0.01801(4)   & 0.43   \\
	& 3  & $2-10$  & 2&        $-$0.01808(6)   &  0.99\\
       &  4  & $2-10$  & 2 &       $-$0.01805(10) &  0.30 \\
       & 5   &$2-10$   &2  &       $-$0.01800(19) & 0.45\\
       &  6  & $2-10$  & 2 &       $-$0.01795(32) & 1.9\F \\
\hline
$\[g^2 B_{y}(0,0) E_{z}(r,t) \]$ 
&2   & $1-10$   & 3  &    {\Fm}0.03618(3) &   2.0\F \\
 &  3   &$1-10$  &  3 &   {\Fm}0.01948(4) &    2.7\F  \\
&   4   &$1-10$  &  3 &   {\Fm}0.01265(10)  &  0.81   \\
 &  5   &$1-10$  &   2 &  {\Fm}0.00917(14)  &  2.0\F   \\
&   6   &$1-10$  &   2 &  {\Fm}0.00795(42)  &  0.88  \\
\hline
$C(t)$ & $r/a$ & Fit range $(t/a)$ & $m_{\rm max}$ & 
$a^3\int_{0}^{\infty} dt \;  C(t)$ & $\chi_{\rm min}^2/N_{\rm df}$ \\
\hline
$\[g^2 B_{x}(0,0) B_{x}(r,t) \]$ 
& 2   &$1-10$  &   3&  {\Fm}0.01648(3)  &  2.2\F\\
&  3   &$1-10$  &3  &   {\Fm}0.00743(3)   &   0.97\\
&   4  &$1-10$  &  3 &  {\Fm}0.00381(3)  &   0.85\\
&   5   &$1-10$  & 2  &  {\Fm}0.00217(3)  &  2.0\F\\
&  6   &$1-10$  &  2  &  {\Fm}0.00128(5)   & 1.8\F\\
\hline
$\[g^2 B_{y}(0,0) B_{y}(r,t) \]$ 
&    2   &$1-10$  &   3&  $-$0.01226(3) &    0.77\\
&    3   &$1-10$  &   3 & $-$0.00443(3)  &   0.40 \\
&    4   &$1-10$  &3  &   $-$0.00155(3)   &  0.37\\
&    5   &$1-10$  &3 &    $-$0.00067(3)   &  0.46\\
&    6   &$1-10$  &  3 &  $-$0.00033(4)  &   0.48\\
\hline
\end{tabular}
\label{tbl:corfit-L20b60}
\end{table}

\begin{table}[hbt]
\caption{%
Fit results of the field strength correlators
at $\beta=6.0$ on the $20^3 40$ lattice.}
\vspace{3mm}
\centering
\begin{tabular}{cccclc}
\hline
$C(t)$ & $r/a$ & Fit range $(t/a)$ & $m_{\rm max}$ & 
$a^2 \int_{0}^{\infty} dt \; t C(t)$ & 
$\chi_{\rm min}^2/N_{\rm df}$ \\
\hline
$\[g^2 B_{y}(0,0) E_{z}(0,t) \]$
&   2  & $2-20$ &   2    & $-$0.01798(13)   & 2.7\\
&    3   & $2-20$ &    2 & $-$0.01809(20) &   2.5\\
&    4  & $2-20$ &   2  &  $-$0.01800(27)  &  2.9\\
&    5  & $2-20$ &  2   &  $-$0.01809(25) &1.4\\
&    6   & $2-20$ &    2 & $-$0.01831(40)  &  1.8\\
&    7   & $2-20$ &  2 &   $-$0.01853(137)  & 2.6\\
\hline
$\[g^2 B_{y}(0,0) E_{z}(r,t) \]$ 
 &  2   & $1-20$ &    3   &{\Fm}0.03618(5)&   1.4\\
 &   3   & $1-20$ &   3   & {\Fm}0.01950(12)&  5.0\\
 &   4   & $1-20$ &    2  &{\Fm}0.01253(18) &  1.9\\
 &   5  & $1-20$ &  2    & {\Fm}0.00920(31)   & 2.4\\
 &   6   & $1-20$ &   2  &{\Fm}0.00703(80)  &  5.7\\
 &   7   & $1-20$ &    2 & {\Fm}0.00457(68)  &  2.6\\
 \hline
 $C(t)$ & $r/a$ & Fit range $(t/a)$ & $m_{\rm max}$ & 
$a^3\int_{0}^{\infty} dt \;  C(t)$ & $\chi_{\rm min}^2/N_{\rm df}$ \\
\hline
$\[g^2 B_{x}(0,0) B_{x}(r,t) \]$ 
& 2     & $1-20$  & 3   &  {\Fm}0.01642(7)  &  1.5\\
&    3   & $1-20$ &  3   & {\Fm}0.00742(5)  &  2.2\\
 &   4   & $1-20$ &  3   & {\Fm}0.00372(7)  &  1.5\\
 &   5   & $1-20$ &  3  & {\Fm}0.00206(6)   & 1.5\\
 &   6   & $1-20$ &  2   & {\Fm}0.00133(16)  &  5.5\\
 &   7   & $1-20$ &  2  & {\Fm}0.00071(18)  & 4.5\\
\hline
$\[g^2 B_{y}(0,0) B_{y}(r,t) \]$ 
&    2   & $1-20$ &   3 &  $-$0.01226(6)&   3.1\\
&    3   & $1-20$ &    3 & $-$0.00442(7) &   2.3\\
&    4   & $1-20$ &    3&  $-$0.00166(7)  &  2.8\\
&    5   & $1-20$ &    3 & $-$0.00069(6) &   2.6\\
&    6   & $1-20$ &    3 & $-$0.00029(9)  &  1.8\\
&    7   & $1-20$ &    3 & $-$0.00030(39)  &  1.9\\
\hline
\end{tabular}
\label{tbl:corfit-L20T40b60}
\end{table}

\begin{table}[hbt]
\caption{%
Fit results of the field strength correlators
at $\beta=6.3$ on the $24^4$ lattice.
($^*$)one of the excitation energies 
in the expansion was fixed so as to make the fit stable.}
\vspace{3mm}
\centering
\begin{tabular}{cccclc}
\hline
$C(t)$ & $r/a$ & Fit range $(t/a)$ & $m_{\rm max}$ & 
$a^2\int_{0}^{\infty} dt \; t C (t)$ & $\chi_{\rm min}^2/N_{\rm df}$ \\
\hline
$\[g^2 B_{y}(0,0) E_{z}(0,t) \]$
 &   2   &$2-12$ &   2 &  $-$0.00980(10)  &    1.1\F\\
  &  3  &$2-12$ &  2 &   $-$0.00872(29)    &  1.2\F\\
  &  4   &$2-12$ &  2 &   $-$0.00768(30)  &   1.4\F\\
 &   5   &$2-12$ & 2 &    $-$0.00727(28)  &   4.4$^{*}$\\
 &   6   &$2-12$ &  2 &   $-$0.00664(110)   &    0.94\\
\hline
$\[g^2 B_{y}(0,0) E_{z}(r,t) \]$ 
 &  2   &$1-12$ &   3 & {\Fm}0.03145(7)    &   0.28\\
  &  3   &$1-12$ &   2& {\Fm}0.01620(16)    &   3.0\F\\
  &  4  &$1-12$ & 2  &  {\Fm}0.01008(37)    &   2.0\F\\
   & 5   &$1-12$ &  2 &  {\Fm}0.00632(43)    &   0.85\\
   & 6   &$1-12$ &   2 &  {\Fm}0.00446(109)  &    2.3\F\\
 \hline
 $C(t)$ & $r/a$ & Fit range $(t/a)$ & $m_{\rm max}$ & 
$a^3\int_{0}^{\infty} dt \;  C(t)$ & $\chi_{\rm min}^2/N_{\rm df}$ \\
\hline
$\[g^2 B_{x}(0,0) B_{x}(r,t) \]$ 
&   2    &$1-12$ &   3 &  {\Fm}0.01452(3)    &  0.49\\
  &  3  &$1-12$ & 3  &   {\Fm}0.00653(3)  &   4.4\F\\
 &   4  &$1-12$ &  2  &  {\Fm}0.00332(6)  &   1.2\F\\
 &   5   &$1-12$ &   2 &  {\Fm}0.00188(8) &     0.87\\
 &   6  &$1-12$ & 1 &   {\Fm}0.00123(12)  &    0.49\\
\hline
$\[g^2 B_{y}(0,0) B_{y}(r,t) \]$ 
&    2  &$1-12$ & 3 &     $-$0.01203(3)  &  0.95\\
  &  3  &$1-12$ &   3 &   $-$0.00436(3)  &   1.4\F\\
 &   4   &$1-12$ &   3 &  $-$0.00168(7) &     1.5\F\\
 &   5   &$1-12$ &   2 &  $-$0.00084(18) &    4.8\F\\
 &   6   &$1-12$ &   1&   $-$0.00046(12)  &   1.8\F\\
\hline
\end{tabular}
\label{tbl:corfit-L24b63}
\end{table}

\clearpage
\subsection{Spin-dependent potentials}

\begin{table}[hbt]
\caption{The spin-dependent potentials at
$\beta=6.0$ on the $20^4$ lattice.}
\vspace*{0.3cm}
\centering
\begin{tabular}{ccccc}
    \hline
    r/a  &$a^2V_{1}'$ &$a^2V_{2}'$ &$a^3V_{3}$ &$a^3V_{4}$ \\
\hline
 2 & $-$0.03603(8)\F  &0.07235(6)\F&0.05749(8)\F& $-$0.01607(13)\\
 3 & $-$0.03616(12) & 0.03896(9)\F &0.02373(8)\F&$-$0.00286(14)\\
 4 &$-$0.03609(20)&0.02531(19)&0.01072(8)\F&{\Fm}0.00143(17)\\
 5 &$-$0.03600(39)& 0.01836(28)&0.00568(9)\F&{\Fm}0.00165(14) \\
 6 & $-$0.03591(65)&0.01590(84)&0.00322(13)&{\Fm}0.00123(20) \\
 \hline
\end{tabular}
\label{tbl:spin-L20b60}
\end{table}

\begin{table}[hbt]
\caption{The spin-dependent potentials at
$\beta=6.0$ on the $20^3 40$ lattice.}
\vspace*{0.3cm}
\centering
\begin{tabular}{ccccc}
\hline
r/a  &$a^2V_{1}'$ &$a^2V_{2}'$ &$a^3V_{3}$ &$a^3V_{4}$ \\
\hline
 2 &$-$0.03596(25)\F&0.07235(11)\F &0.05736(19)\F &$-$0.01620(27)\F \\
 3 &$-$0.03618(41)\F& 0.03900(25)\F &0.02368(20)\F &$-$0.00284(29)\F \\
 4 &$-$0.03600(54)\F& 0.02507(36)\F &0.01075(21)\F & {\Fm}0.00081(30)\F \\
 5 &$-$0.03619(50)\F&0.01840(61)\F & 0.00549(18)\F &{\Fm}0.00138(27)\F \\
 6 &$-$0.03662(81)\F&0.01405(160) &0.00323(37)\F &{\Fm}0.00152(47)\F \\
 7 &$-$0.03706(274)&0.00914(136) &0.00203(103)&{\Fm}0.00021(137)\\
\hline
\end{tabular}
\label{tbl:spin-L20T40b60}
\end{table}

\begin{table}[hbt]
\caption{The spin-dependent potentials at
$\beta=6.3$ on the $24^4$ lattice.}
\vspace*{0.3cm}
\centering
\begin{tabular}{ccccc}
\hline
r/a  &$a^2V_{1}'$ &$a^2V_{2}'$ &$a^3V_{3}$ &$a^3V_{4}$ \\
\hline
 2 &$-$0.01960(19)\F& 0.06290(15)\F& 0.05310(9)\F&  $-$0.01910(15)\\
 3 &$-$0.01744(58)\F& 0.03240(31)\F& 0.02178(9)\F&  $-$0.00439(14) \\
 4 &$-$0.01539(61)\F& 0.02017(74)\F& 0.01000(16)&   $-$0.00009(31) \\
 5& $-$0.01443(79)\F& 0.01265(87)\F& 0.00546(41)&  {\Fm}0.00038(70) \\
 6 &$-$0.01329(221) &0.00893(217) & 0.00339(37)&{\Fm}0.00057(52) \\
\hline
\end{tabular}
\label{tbl:spin-L24b63}
\end{table}

\clearpage

\end{document}